\documentclass[twocolumn]{aastex62}
\usepackage{url,graphics,array,amsmath,amssymb,ctable,enumitem}
\usepackage{booktabs}
\usepackage[flushleft]{threeparttable}

\newcommand\clearrow{\global\let\rowmac\relax}

\newcommand{\zreion}{z_{\mathrm{re}}}
\newcommand{\HI}{\mathrm{H\,I}}
\newcommand{\HII}{\mathrm{H\,II}}

\newcommand{\Msun}{M_{\odot}}
\newcommand{\Mpc}{\mathrm{Mpc}}

\newcommand{\Treion}{T_{\mathrm{reion}}}

\newcommand{\MFP}{\lambda_{\mathrm{mfp}}^{912}}
\newcommand{\Lbox}{L_{\mathrm{box}}}

\newcommand{\vIF}{v_{\mathrm{IF}}}

\shorttitle{Hydrodynamic Response to Reionization}
\shortauthors{D'Aloisio et al.}

\begin{document}

\title{Hydrodynamic Response of the Intergalactic Medium to Reionization}

\author{Anson D'Aloisio}
\email{ansond@ucr.edu}
\affil{Department of Physics \& Astronomy, University of California, Riverside, CA 92521, USA}

\author{Matthew McQuinn}
\affil{Astronomy Department, University of Washington, Seattle, WA 98195, USA}

\author{Hy Trac}
\affiliation{McWilliams Center for Cosmology, Department of Physics, Carnegie Mellon University, Pittsburgh, PA 15213, USA}

\author{Christopher Cain}
\affiliation{Department of Physics \& Astronomy, University of California, Riverside, CA 92521, USA}

\author{Andrei Mesinger}
\affiliation{Scuola Normale Superiore, 56126 Pisa, PI, Italy}

\begin{abstract}
The intergalactic medium is expected to clump on scales down to $10^4-10^8$ M$_{\odot}$ before the onset of reionization. The impact of these small-scale structures on reionization is poorly understood despite the modern understanding that gas clumpiness limits the growth of $\HII$ regions. We use a suite of radiation-hydrodynamics simulations that capture the $\sim 10^4\Msun$ Jeans mass of unheated gas to study density fluctuations during reionization.   Our simulations track the complex ionization and hydrodynamical response of gas in the wake of ionization fronts. The clumping factor of ionized gas (proportional to the recombination rate) rises to a peak value of $5-20$ approximately $\Delta t = 10$ Myr after ionization front passage, depending on the incident intensity, redshift, and degree to which the gas had been pre-heated by the first X-ray sources.  The clumping factor reaches its relaxed value of $\approx 3$ by $\Delta t = 300$ Myr. The mean free path of Lyman-limit photons evolves in unison, being up to several times shorter in un-relaxed, recently reionized regions compared to those that were reionized much earlier. Assessing the impact of this response on the global reionizaton process, we find that un-relaxed gaseous structures boost the total number of recombinations by $\approx 50$ \% and lead to spatial fluctuations in the mean free path that persist appreciably for several hundred million years after the completion of reionization.          
\end{abstract}


\section{Introduction}

A consistent picture of reionizaton has begun to emerge from observations.  Measurements of the Cosmic Microwave Background (CMB) anisotropies by the Planck collaboration imply a reionization midpoint of $z=7.7 \pm 0.7$ \citep{2018arXiv180706209P}.  This is supported by a number of independent astrophysical probes.  Evidence of damping wing absorption in two of the three known $z>7$ quasars \citep{mortlock11,2017MNRAS.466.4239G,2018Natur.553..473B, 2018ApJ...864..142D}, the steep decline in the fraction of galaxies that show Ly$\alpha$ emission at $z>6$ \citep[e.g.][]{2006ApJ...648....7K, schenker12,2014ApJ...793..113P, 2015MNRAS.446..566M, ono12}, and the lack of transmission in the $z>6.2$ Ly$\alpha$ forest \citep{fan06, 2019ApJ...884...30W}, all support the view that much of the intergalactic medium (IGM) was neutral at these redshifts.  Together with a direct census of the galaxy and quasar populations \citep[e.g.][]{2015ApJ...803...34B, 2019ApJ...879...36F, 2019MNRAS.488.1035K}, these observations favor a scenario in which the bulk of reionization occurred between $z=5.5-10$ and was driven mostly by the earliest galaxies.

The reionization process is shaped by both the sources and the small-scale density structure of the IGM.  Of particular importance are the so-called {\it sinks} of ionizing photons, or self-shielding systems -- intergalactic absorbers with photoelectic optical depth to 1 Ry photons $>$ 1. (In what follows, we will use the terms sink and self-shielding region interchangeably.) Numerous papers have been written on how reionization is affected by the properties of the sources \citep[see][and references therein]{2016ARA&A..54..313M}.  Comparatively little effort has gone into understanding the impact of small-scale structure.  That is the main focus of this paper.

\begin{figure*}
\includegraphics[width=18cm]{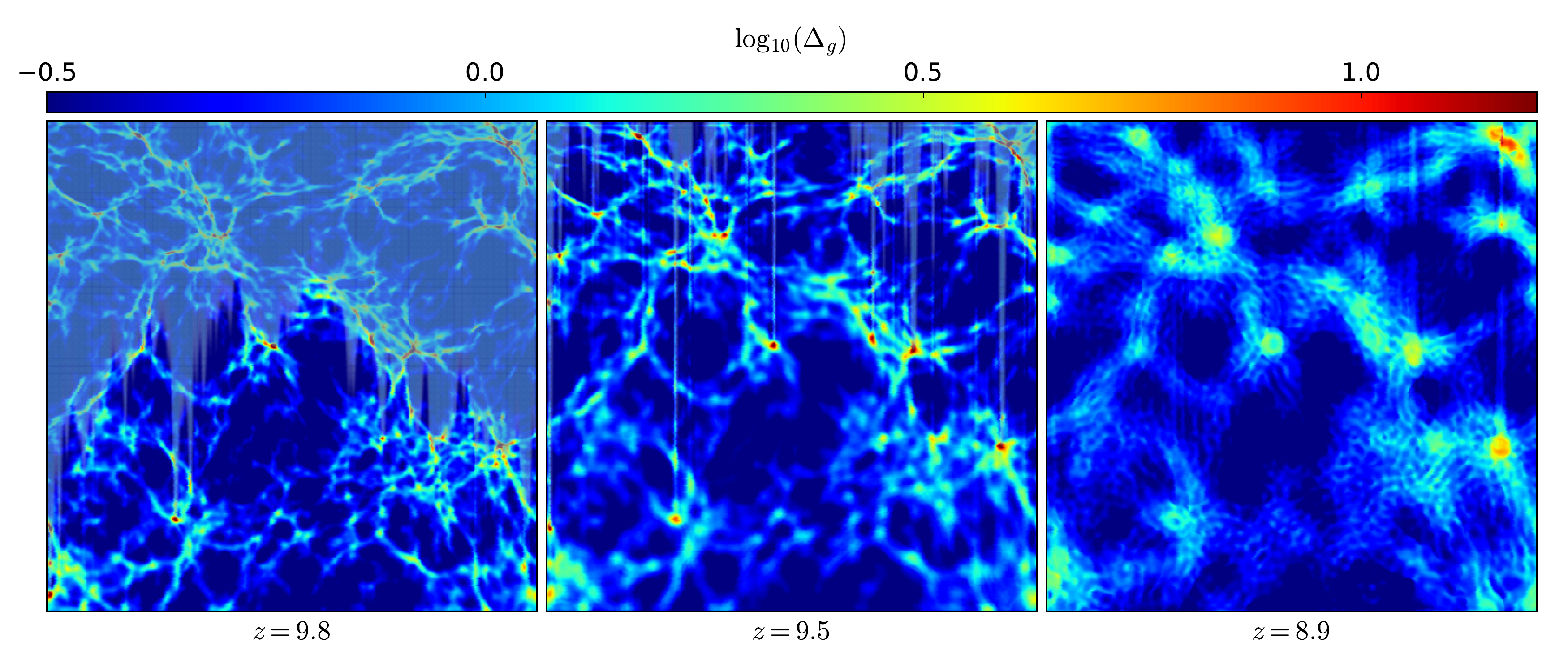}
\caption{ Hydrodynamic response of the IGM during reionization. The panels show three snapshots from a ray-tracing radiation-hydrodynamics simulation in which plane parallel radiation is turned on at $z=10$. Shaded regions are neutral and the sharp boundary in the left panel corresponds to the ionization front.   Each panel is $0.5 h^{-1}$ Mpc on a side and the simulation uses an Eulerian grid with $N=512^3$ cells (the code will be described in \S \ref{sec:methods}).  The gas density, $\Delta_g$, is given in units of the cosmic mean. At early times (left panel), the gas clumps on scales down to the $\sim 10^4$ M$_{\odot}$ Jeans mass of the unheated IGM.  The supersonic I-front sweeps through the volume ahead of the hydrodynamic response, getting trapped in some locations by self-shielding density peaks (middle panel).  Over time the density structure is smoothed out as the gas responds to the impulsive heating by the ionization front (right panel).   }
\label{fig:intro}
\end{figure*}

It is important to model the effects of small-scale structure accurately for several reasons. The recombination rate of the IGM, which is proportional to the variance of density fluctuations, sets the total number of ionizing photons required to reionize the Universe and to maintain it thereafter. To understand what reionization reveals about the sources driving it, we must also understand the small-scale gas distribution. Secondly, small-scale structure impacts the morphology of $\HII$ regions during reionization. The growth of an ionized bubble will be slowed by absorptions once it is larger than the mean free path of ionizing photons to be absorbed within a sink \citep{2005MNRAS.363.1031F, 2007MNRAS.377.1043M, 2012ApJ...747..126A, 2014MNRAS.440.1662S}. In addition, self-shielding could produce islands of neutral gas inside of $\HII$ regions \citep{2009MNRAS.394..960C, 2011MNRAS.411..289C}.     Aside from the large-scale CMB anisotropies, all reionization observables are sensitive to its morphology and, thus, are impacted to some extent by the sinks.

One complication to modelling these effects is that the density structure of the IGM evolves significantly in response to the heating from reionization. To illustrate this point, in Fig. \ref{fig:intro} we show snapshots from a radiative hydrodynamics simulation of an ionization front (I-front) sweeping through the IGM.  The light blue shaded regions correspond to neutral gas.  Immediately after a region has been reionized, the gas clumps on mass scales as small as $10^4~M_\odot$, corresponding to $\sim 1$ physical kpc, the Jeans scale of the pre-reionization gas.  Afterwards, the gas begins to relax dynamically and redistribute in response to the photoheating from reionization, evacuating from small potential wells over a sound crossing time \citep{2004MNRAS.348..753S, 2005MNRAS.361..405I}.  The relaxation can take hundreds of millions of years, ultimately smoothing the gas on $100~$kpc scales and over masses of $10^9~M_\odot$ \citep{2001ApJ...559..507S}. The right panel in Fig. \ref{fig:intro} shows a more relaxed state 80 million years after the initial ionization. Simulating these processes requires high-resolution cosmological hydrodynamics (hydro) coupled to radiative transfer (RT).

In this paper we use a suite of radiation-hydrodynamics simulations to study small-scale structure and its impact on reionization.  The central questions that we aim to address are: (1) How do the opacity and recombination rate of the IGM evolve in response to reionization?; (2) How do they depend on environmental factors such as the local intensity of the ionizing background and density?; (3) What role does the relaxation of the IGM play in the global reionization process?

The work presented here builds on previous numerical studies, which fall mainly into two classes.  The first class uses cosmological radiative-transfer (RT) simulations of reionization to explore the role of small-scale structure \citep[e.g.][]{2014ApJ...789..149S, 2015ApJ...810..154K, 2018MNRAS.478.1065C, 2018MNRAS.478.5123R}.  These studies include prescriptions for modelling galaxy formation at varying levels of complexity.  Recently, it has become possible to run coupled hydro$+$RT simulations of reionization in boxes up to $L_{\rm box} = 100 h^{-1}$ Mpc \citep[e.g.][]{2014ApJ...793...29G, 2016MNRAS.463.1462O, 2019ApJ...870...18D}.  The main disadvantages of this class are: (1) At present, it is computationally infeasible to capture the large-scale structure of reionization, which requires box sizes with $L > 100$ Mpc \citep{2014MNRAS.439..725I}, while also resolving the kpc-scale size of unrelaxed gas clumps; (2) The complexity of the simulations render it difficult to interpret how the mean free path and gas clumping are affected by the various physical processes at play. 

The second class of study avoids these issues by running simplified simulations in smaller volumes (and therefore at higher resolutions).  In contrast to the full reionization simulations, this class does not attempt to model the full complexity of the source population, but instead applies a ``controlled" radiation field to the gas.  Earlier studies applied a uniform radiation field (no RT) with a simple density threshold criterion to approximate the effects of self-shielding \citep[e.g.][]{2000ApJ...530....1M, 2009MNRAS.394.1812P, 2012ApJ...747..100S, 2012arXiv1209.2489F}.  These provide a rough estimate for the clumping factor, but their results are generally limited to the relaxed limit, and are accurate at the $\sim 50\%$ level owing to the simplistic modeling of self-shielding.  Other studies have attempted to account more realistically for self-shielding by post-processing cosmological hydro simulations with RT.  This approach is more accurate but has only been applied in the fully unrelaxed limit \citep{2013ApJ...763..146E} or the fully relaxed limit \citep{2011ApJ...743...82M, 2011ApJ...737L..37A, 2013MNRAS.430.2427R, 2018MNRAS.478.1065C}.\footnote{The relaxed-limit calculations post-process simulations in which the gas had been heated by a uniform ultraviolet background at a much earlier time.  The first three studies referenced also focused on the post-reionization IGM.}  We note that the unrelaxed limit yields recombination rates that are a factor of $5-10$ larger than the relaxed case \citep{2013ApJ...763..146E}, motivating our work to quantify the transition between these regimes, which we show takes hundreds of millions of years --  a substantial fraction of the duration of reionization. More recently, \citet{2016ApJ...831...86P} explored the evolution of small-scale structure through this transition using a suite of small-volume hydro simulations coupled to an approximate model for RT (most of their runs use $L_{\rm box} = 200 h^{-1}$ kpc).

The current paper fits into the second class and is most similar to the study of \citet{2016ApJ...831...86P}.  Our work includes major improvements over their pioneering study.  The simulations in \citet{2016ApJ...831...86P} employed an approximate scheme for the RT based on local gradients in density that loosely mimics the full RT performed here. Furthermore, structure formation is strongly suppressed in such small boxes.  We are able to achieve comparable resolution to \citet{2016ApJ...831...86P} with $\approx 130$ times the volume.  We also use DC modes \citep[e.g.][]{2011ApJS..194...46G} and a new method of averaging over them based on Gauss-Hermite Quadrature to model the effects of larger structures that our simulations miss.

The remainder of this paper is organized as follows. In \S \ref{sec:scales}, we provide intuition for the characteristic scales at play in the IGM during reionization. In \S \ref{sec:methods} we describe our simulation setup and analysis.  In \S \ref{sec:results} we present our main results and in \S \ref{sec:implications} we discuss their implications for the global reionization process. Finally, in \S \ref{sec:conclusion} we offer concluding remarks.  Throughout this paper we adopt a flat $\Lambda$CDM cosmology with $\Omega_m = 0.31$, $\Omega_b=0.048$, $H_0=100 h$ km s$^{-1}$ Mpc$^{-1}$, with $h=0.68$, $\sigma_8=0.82$, $n_s = 0.9667$, and a hydrogen mass fraction of $X_{\rm Hy} = 0.7547$, consistent with the latest Planck measurements \citep{2018arXiv180706209P}.  Unless otherwise noted, all distances are reported in comoving units.

\section{Characteristic scales}
\label{sec:scales}

We begin by considering the relevant distance, time, and velocity scales that frame the photoionization and photoheating processes under study.  There are two velocity scales of importance during reionization: the speed of I-fronts and the speed of sound.  The former is determined by the number flux of ionizing photons, $F$, at the front boundary according to the condition $\vIF n_{\mathrm{H}} = F$, where $\vIF$ is the front speed and $n_{\mathrm{H}}$ is the proper hydrogen density.\footnote{This expression applies in the non-relativistic limit, which is a good approximation for the galactic sources that likely drove reionization.}  We can write the proper I-front speed in terms of the hydrogen photoionization rate\footnote{In detail, to derive  equation~\ref{eqn:vIF} we adopt a spectrum of the form $I_\nu \propto (\nu/\nu_{\HI})^{-\alpha}$ with a sharp cutoff at 4 Ry, where $I_\nu$ is the specific intensity and $h \nu_{\HI} = 13.6$ eV.  We also assume a photoionization cross section of $\sigma_{\HI} \propto (\nu/\nu_{\HI})^{-2.8}$, which is an excellent approximation at the energies of interest.}, 
\begin{equation}
\vIF = \left(3.8\times 10^3 \frac{\mathrm{km}}{\mathrm{s}}\right)~\Delta_g^{-1} \left( \frac{1+z}{8} \right)^{-3} \left(\frac{\Gamma_{-12}}{0.1} \right).
\label{eqn:vIF}
\end{equation}
Here, $\Delta_g$ is the gas density in units of the cosmic mean and $\Gamma_{-12}$ is the photoionization rate in units of $10^{-12}$ s$^{-1}$, with the $z\sim 5.5$ Ly$\alpha$ forest suggesting $\langle \Gamma_{-12} \rangle \approx 0.3-0.5$ \citep{2018MNRAS.473..560D,2019MNRAS.490.3177W} and {\cal O}(1) spatial fluctuations about this value \citep{2009MNRAS.400.1461M, 2015arXiv150907131D}.  Thus, I-front speeds are typically in the range $\sim 10^3 -10^4$ proper km s$^{-1}$, with the faster speeds occurring near the end of reionization \citep{2019ApJ...874..154D,2019A&A...622A.142D}.  

For the bulk of the IGM during reionization, the I-front speeds are much larger than the sound speed of the gas, 
\begin{equation}
c_s \approx \left( 22 ~\frac{\mathrm{km}}{\mathrm{s}}\right) \left( \frac{T}{20,000\;\mathrm{K}}\right)^{1/2},
\label{eq:soundspeed}
\end{equation}
where $T$ is the gas temperature. (Here we have assumed an ideal gas of fully ionized hydrogen and singly ionized helium of primordial composition and adiabatic index $\gamma = 5/3$.). An exception to this is the high density ($\Delta_g \gtrsim 100$ ) gas surrounding halos, where I-fronts can slow down to speeds of order $c_s$ and below.

The characteristic coherence length of overdense gas is the distance a sound wave can travel in a free fall (or dynamical) time -- the Jeans length -- which in comoving units is
\begin{equation}
    L_J = 25~h^{-1}\mathrm{kpc}~\left(\frac{8}{1+z} \right)^{1/2}\left(\frac{T}{10^4~\mathrm{K}} \right)^{1/2} \left(\frac{10}{\Delta_g}\right)^{1/2}. 
    \label{eq:LJ}
\end{equation}
Below, we will see that the actual sizes of self-shielding regions can be considerably larger than this during reionization.\footnote{However, Eqn.~\ref{eq:LJ} has been shown to hold at at a surprisingly quantitatively level for overdense systems after reionization \citep[see e.g.][]{2001ApJ...559..507S, 2011ApJ...743...82M, 2018MNRAS.478.5123R}.}  Before reionization, the intergalactic gas temperatures are expected to be in the range $T\sim 10-1,000$ K \citep{2006MNRAS.371..867F, 2014Natur.506..197F}.  When an I-front sweeps through a region, it impulsively heats the gas to temperatures $20,000 - 30,000$ K \citep{2019ApJ...874..154D}, such that $L_J$ has to adjust by a factor of $5-50$.  A rough estimate for the timescale of this relaxation is the sound crossing time of the absorption systems with size $L_J$, i.e. the dynamical time,
\begin{equation}
t_{\mathrm{dyn}} = \frac{1}{\sqrt{4 \pi G \rho_m}} = 300~\mathrm{Myr}~\left( \frac{10}{\Delta_g}\right)^{1/2} \left(\frac{1+z}{8} \right)^{-3/2} ,
\end{equation}
where $G$ is Newton's gravitational constant, and $\rho_m$ is the total matter density.\footnote{At lower densities, the characteristic time is $H^{-1} \approx 1~\mathrm{Gyr}~((1+z)/8)^{-3/2}$, where $H$ is the Hubble parameter.} These arguments imply that over-densities $\Delta_g \sim 10$, which are thought to contribute substantially to the Lyman-limit opacity, have a relaxation timescale of a few hundred million years. 

Not only will these velocity, distance, and time scales prove helpful for interpreting our results; as we will see in the next section, our simulation setup was optimized according to them.

\section{Numerical Methodology}
\label{sec:methods}

\subsection{Radiative Hydrodynamics Simulations}

We ran a suite of high-resolution hydro$+$RT simulations using a modified version of the ray-tracing code of \citet{2008ApJ...689L..81T} \citep[see also][]{2004NewA....9..443T,2007ApJ...671....1T}.  Our simulations can be thought of as a set of controlled numerical experiments in which we send I-fronts through a small patch of the IGM and track how the gas responds.  In this spirit, we do not implement the complex physical processes associated with galaxy formation. The hydro module does, however, track all of the relevant heating and cooling processes for gas of primordial composition (see description in \citealt{2018MNRAS.473..560D}). We initialized our simulations at $z=300$ using first-order perturbation theory and separate transfer functions for the baryonic and dark matter obtained from CAMB \citep{Lewis:1999bs}.\footnote{We checked that our simulations reproduce linear theory.} Unless otherwise stated, the gas evolves adiabatically after decoupling with the CMB (at $z\approx 150$), until the ionizing radiation turns on.        

The multi-frequency ray tracing is implemented in a plane parallel geometry on a uniform grid with 5 frequency bins spanning $1$Ry to $4$Ry.  We adopted a power-law spectrum with specific intensity $J_\nu \propto \nu^{-1.5}$ (where $\nu$ is frequency), which roughly mimics the un-absorbed spectral energy distributions of stars and quasars.  However, as the heating is only weakly dependent on the spectral index, we expect our results to be insensitive to this choice  \citep{2016MNRAS.456...47M, 2019ApJ...874..154D}.  We set the frequency binning such that the number density of photons is the same for all bins.   The RT grid's resolution is matched to that of the hydro solver, with the option of sending rays from any of the 6 directions simultaneously.  For all simulations described herein, we chose to send rays along two orthogonal directions.  This configuration yields a planar propagation of the I-front while at the same time avoiding unrealistic shadowing effects in a single direction. Simulations of reionization suggest that radiation from multiple directions (and many sources) contributes to the flux at the I-front. 

To reduce computational costs, we implemented an adaptive speed of light approximation motivated by the characteristic velocity scales in \S \ref{sec:scales}. We set the speed of light to $c_{\rm sim} =  0.1 c$ (where $c=3\times10^5$ km/s) until the volume-weighted mean $\HI$ fraction of the box dropped below 1\%.  Below this threshold we set $c_{\rm sim} =  0.01 c$.  Our adaptive algorithm ensures that $c_{\rm sim}$ is much greater than the I-front speeds while the box is being ionized, and the sound speed thereafter. We have tested our algorithm against a simulation in which $c_{\rm sim} =  c$. In Appendix \ref{sec:convergence} we show that the adaptive algorithm is accurate to better than $10$~\%.

Our fiducial box sizes and resolutions were $L_{\rm box} = 1.024h^{-1} ~\Mpc$ and $N \equiv N_{\mathrm{gas}} = N_{\mathrm{dm}}=N_{\mathrm{rt}}=1024^3$, respectively (where $N_{\mathrm{gas}}$, $N_{\mathrm{dm}}$, and $N_{\mathrm{rt}}$ are the hydro solver grid, dark matter particle, and RT grid numbers, respectively).  Note that this corresponds to a gas resolution of $\Delta x = 1 h^{-1}~\mathrm{kpc}$, roughly the Jeans scale of the adiabatically evolving gas prior to ionization, and a dark matter particle mass of $105~M_{\odot}$.  In Appendix \ref{sec:convergence} we present numerical convergence tests with respect to resolution. Our simulations are well converged for $\Delta t > 10$ Myr after ionization.  At earlier times they are converged at only the factor of 2 level, suggesting that structures smaller in size than $\Delta x= 10h^{-1}$ kpc contribute to the clumping at these times. In what follows, we will find that uncertainties in the degree of pre-heating by the first X-ray sources lead to a similar variations at $\Delta t < 10$ Myr.

We aim to study the evolution of small-scale structure after it has been impulsively heated by passing I-fronts. If we had placed the ray sources at the boundaries of the box, as we did for the illustrative example shown in Fig~\ref{fig:intro}, the I-fronts would have taken several tens of millions of years to traverse the box.  This would complicate our interpretation of the evolution because the difference in reionization times for gas parcels separated by roughly one box length would have been a significant fraction of the relaxation time scale.  Another issue is that the shadowing effects of self-shielding systems would have caused the intensity of the ionizing radiation to be significantly lower behind them, potentially resulting in a strong gradient across the box.  We would like to control the intensity to quantify how relaxation depends on the local radiation background. 

To mitigate these problem, we divide our simulation box into $32^3$ cubical ``domains" with side lengths $L_{\rm dom} = 32~h^{-1}$kpc and we send rays from a grid of sources placed on the boundaries of the domains.  The photon number densities do not dilute with cosmological expansion and rays are deleted once they reach the ends of their respective domains.   This way, all of the gas in the box is reionized at nearly the same time, and is exposed to a fixed radiation intensity at all times.  We omit source cells for all calculated quantities discussed in this paper.  We note that our domain sizes are somewhat smaller than the $\sim 50 h^{-1}$ kpc-wide zones around I-fronts within which photoheating and line cooling set the post-I-front temperature, $T_{\rm reion}$ \citep{2019ApJ...874..154D}.  In spite of this, we find that $\Treion \approx 20,000$ K in our simulations, consistent with the range of values reported by \citet{2019ApJ...874..154D}.  One might also worry that the domain structure ionizes over-dense gas that would have remained neutral otherwise, potentially resulting in a spuriously high recombination rate.  In the Appendix we test for such an effect using larger domains.  The test simultaneously demonstrates the need for small domains, and shows that our results are insensitive to the domain size once the I-fronts have traversed the box.

We parameterize the radiation field with $\Gamma_{-12}$ and $\zreion$. The former is the hydrogen photoionization rate in our source cells (at the boundaries of the domains), expressed in units of $10^{-12}$ s$^{-1}$.  The latter is the redshift at which the radiation turns on. Thanks to our domain setup, we may also take this to be the redshift at which the entire simulation domain is ``reionized'' -- i.e. a cosmological I-front has passed and neutral gas is confined to dense self-shielding structures. 

Table \ref{tab:sims} lists all of the runs performed for this paper. (In the next section we will describe the box-scale density parameter, $\delta/\sigma$.)  Note that our runs generally consist of permutations of $\zreion=6$, 8, and 12, and $\Gamma_{-12}=0.3$ and 3.0. We have also performed one run with $\Gamma_{-12}=0.1$. In addition, one of our runs quantifies the impact of pre-heating of the gas by the first astrophysical X-ray sources during cosmic dawn. For this we simply impose a temperature floor of $T_i = 1,000$ K starting at $z=20$.  This is in the hotter range of expected temperatures, and existing models generally predict the bulk of the X-ray heating to occur closer to the time of ionization \citep{2006MNRAS.371..867F, 2014Natur.506..197F}, which would result in less relaxation compared to our artificial temperature floor.  In this sense, the run provides a crude upper limit for the possible effects.\footnote{A larger effect potentially would be if the dark matter is warm or fuzzy, which would erase the smallest fluctuations.  Constraints from the Ly$\alpha$ forest suggest that the dark matter cannot be so warm to erase clumps of mass $\sim 10^7-10^8M_\odot$ \citep[e.g.][]{2013PhRvD..88d3502V, 2017PhRvD..96b3522I}.  A cosmology that saturates this bound would erase much of the structure that contributes to the relaxation.}   Lastly, we have run a set of 10 simulations for convergence tests, which are described in Appendix \ref{sec:convergence}.

Our controlled ``numerical experiment" setup fixes $\Gamma_{-12}$ in time. One might expect that the local $\Gamma_{-12}$ of a recently reionized patch increases with time as $\HII$ bubbles grow to encompass more sources. \citet{2014MNRAS.440.1662S} found, however, that the evolution of $\Gamma_{-12}$ is reasonably flat on average.  The expanding number of sources within a given bubble is tempered by the increasing cost to reionize a larger volume (and to maintain it thereafter).    Moreover, we argue that it is the {\it initial} radiation intensity, i.e. what the gas is exposed to within the first $\sim10$ Myr of ionization, that dominates the evolution of the density structure during relaxation.  If $\Gamma_{-12}$ evolves with time, the gas will subsequently adjust over a longer time scale, but this relaxation will be sub-dominant to that from the impulsive heating of I-fronts.

\begin{table}
    \caption{Radiative hydrodynamics simulations run for this study.}
     \begin{tabular}{rccccc}
             \toprule
        $\zreion$ & $\Gamma_{-12}^{*}$ & $\delta/\sigma$ & $\Lbox^\dagger$ & $N$  & comment \\
        \midrule
            \multicolumn{2}{c}{\footnotesize{\it Production runs}} & & &    \\
     \midrule
     12 & 0.3 & $0$  & $1.024$   & $1024^3$ &  \\
     12 & 3.0 & $0$  & $1.024$   & $1024^3$ &  \\
     8 & 0.1 & $0$  & $1.024$   & $1024^3$ &   \\ 
   
     8 & 0.3 & $0$  & $1.024$   & $1024^3$ &  \\ 
     8 & 0.3 & $0$  & $1.024$   & $1024^3$ & $T_i = 1,000$K \\ 
     8 & 0.3 & $0$  & $2.048$   & $1024^3$ &  \\
    8 & 0.3 & $+\sqrt{3}$  & $1.024$   & $1024^3$ &  \\ 
     8 & 0.3 & $-\sqrt{3}$  & $1.024$   & $1024^3$ &  \\ 
     8 & 3 & 0  & $1.024$  & $1024^3$  & \\ 
     8 & 3 & $+\sqrt{3}$  & $1.024$   & $1024^3$ &  \\ 
     8 & 3 & $-\sqrt{3}$  & $1.024$   & $1024^3$ &  \\ 
     6 & 0.3 & 0  & $1.024$   & $1024^3$ &  \\    
     6 & 0.3 & $+\sqrt{3}$  & $1.024$   & $1024^3$ &  \\ 
     6 & 0.3 & $-\sqrt{3}$  & $1.024$   & $1024^3$ &  \\ 
     6 & 3 & 0  & $1.024$  & $1024^3$  &  \\ 
     \midrule
     \multicolumn{2}{c}{\footnotesize{\it Convergence runs}} & & &    \\
     \midrule
      8 & 0.3 & $0$  & $0.256$   & $256^3$ & $c_{\rm sim}= c$  \\  
       8 & 0.3 & $0$  & $0.256$   & $256^3$ & $c_{\rm sim}= 0.1c$  \\  
        8 & 0.3 & $0$  & $0.256$   & $256^3$ & $c_{\rm sim}= 0.01c$  \\
        8 & 0.3 & $0$  & $0.256$   & $256^3$ &   \\
         8 & 0.3 & $0$  & $0.256$   & $1024^3$ &   \\ 
         8 & 0.3 & $0$  & $0.256$   & $512^3$ &   \\  
         8 & 0.3 & $0$  & $0.256$   & $128^3$ &  \\  
         8 & 0.3 & $0$  & $0.256$   & $64^3$ &  \\
         8 & 0.3 & $+\sqrt{3}$  & $0.256$   & $256^3$ &  \\  
         8 & 0.3 & $+\sqrt{3}$  & $0.256$   & $256^3$ & $N_{\rm dom}=1$  \\
        \bottomrule
     \end{tabular}
    \begin{tablenotes}
      \small
      \item * In units of $10^{-12}~s^{-1}$
      \item $\dagger$ In units of comoving $h^{-1}$ Mpc
    \end{tablenotes}
    \label{tab:sims}
\end{table}

\subsection{DC Modes}
 
Our box sizes strike a balance between computational cost and achieving the high resolutions necessary to capture small-scale structure. However, the variance of linear density fluctuations in spheres of radius $1(2)~h^{-1}\Mpc$ is already $0.6(0.4)$ by a redshift of $z=5.5$.  If our suite consisted only of simulations normalized to the cosmic mean density, we would miss effects from the large density variations on our $L=1.024~h^{-1}$Mpc box scale.  To model these effects, we have modified the simulation code to add a uniform background density -- termed a ``DC mode" -- to the box.  We follow the approach of \citet{2011ApJS..194...46G}, except that our implementation accounts for the full non-linear evolution of the DC mode\footnote{An alternative method exploits the fact that adding a uniform density to a simulation is equivalent to modifying the cosmological parameters, including a non-zero spatial curvature.  This approach has been termed the ``separate universe" approach \citep{2005ApJ...634..728S, 2015MNRAS.448L..11W}.}. This is important because fluctuations on the box scale are already in the non-linear regime by $z= 5.5$. We parameterize the environmental density of our boxes with $\delta/\sigma$, the ratio of the mean linear perturbation theory density contrast on the box scale (the box's `zero mode'), to the standard deviation of linear fluctuations on this scale (see Table 1). Note that $\delta/\sigma$ does not vary with time.

The method of DC modes serves a twofold purpose for us: (1) It allows us to test how quantities such as the mean free path and recombination rate scale with the environmental density; (2) By averaging the results from our simulations appropriately, we can effectively sample the full distribution of IGM densities. To this end, we developed a novel method of averaging over DC mode simulations based on Gauss-Hermite Quadrature, which we describe in Appendix~\ref{DCmodes}. For three samples of the box-scale density, the accuracy of the average is maximized by choosing $\delta/\sigma=0,\pm \sqrt{3}$ based the roots of the Hermite polynomial, $H_3(x)$.\footnote{For reference, $\delta/\sigma = \pm \sqrt{3}$ corresponds to $\delta_0 = \pm 5$, where $\delta_0$ is linearly extrapolated to the present day.} The over-dense runs with $\delta/\sigma=\sqrt{3}$ would reach turnaround at $z\approx 5.0$, so we terminated them before reaching this point.

\section{Results}
\label{sec:results}
 
\begin{figure*}
\includegraphics[width=18cm]{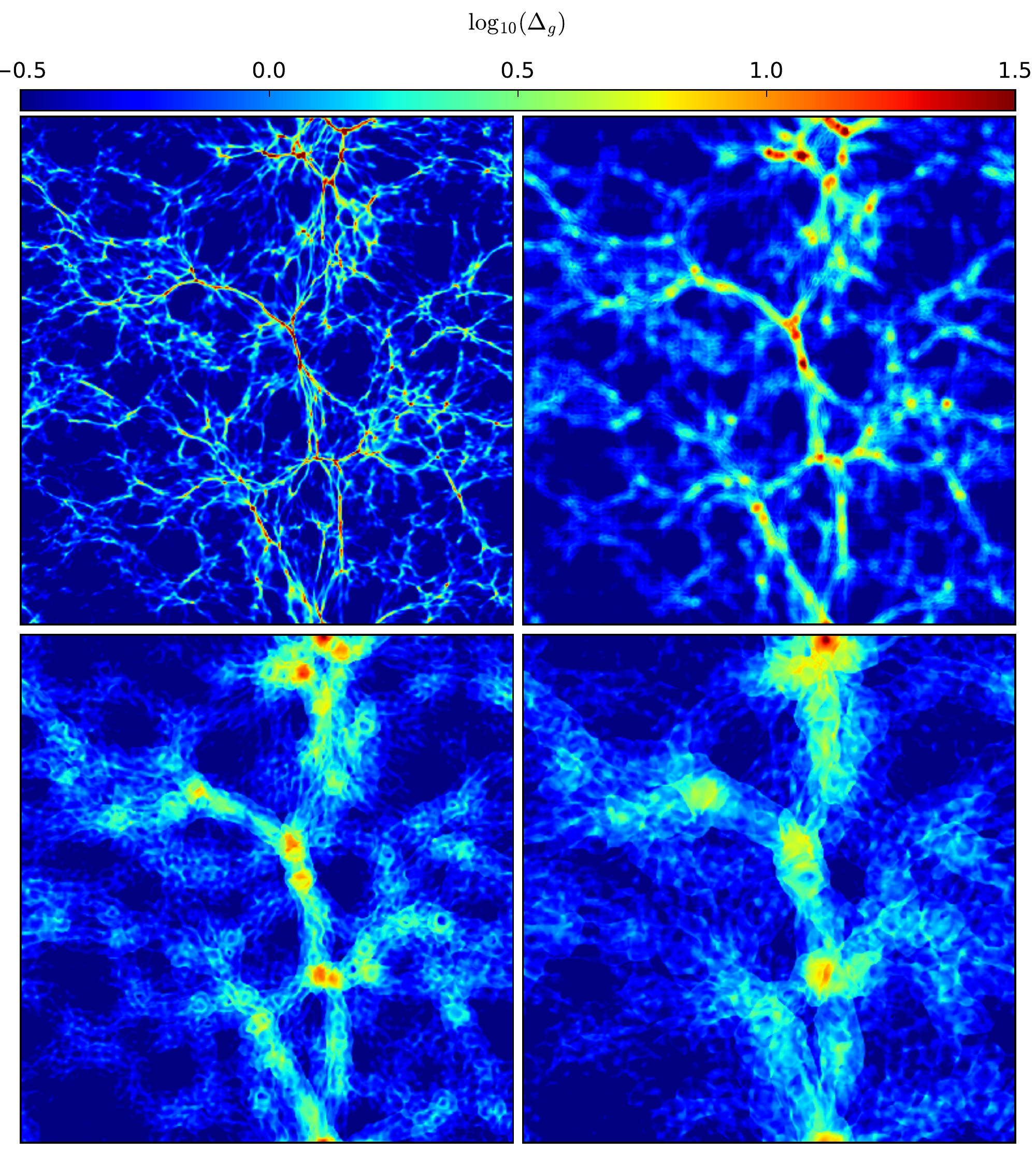} 
\caption{Relaxation of the IGM at $\Delta t= 10$ (top-left), 60 (top-right), 150 (bottom-left), and 300 (bottom-right) Myr after I-fronts have crossed the simulation box.  The panels show slices, two cells thick, through our simulation with $\Gamma_{-12} = 0.3$ and $\zreion=8$, color-coded by the log of the density.  Each panel is $L_{\rm box} = 1h^{-1}$ Mpc across.  At $\Delta t= 10$ Myr, insufficient time has transpired to erase the smallest gaseous structures that were present in the unheated IGM.   By $\Delta t= 300$ Myr the gas has largely reached its relaxed state, with weak shock fronts and interference patterns from the sound waves that drove this relaxation strikingly visible. The sound waves are initialized at the time of ionization, leading to a coherent acoustic scale in the gas density that can be seen in the bottom two panels. }
\label{fig:gas_density}
\end{figure*}

\begin{figure*}
\includegraphics[width=18cm]{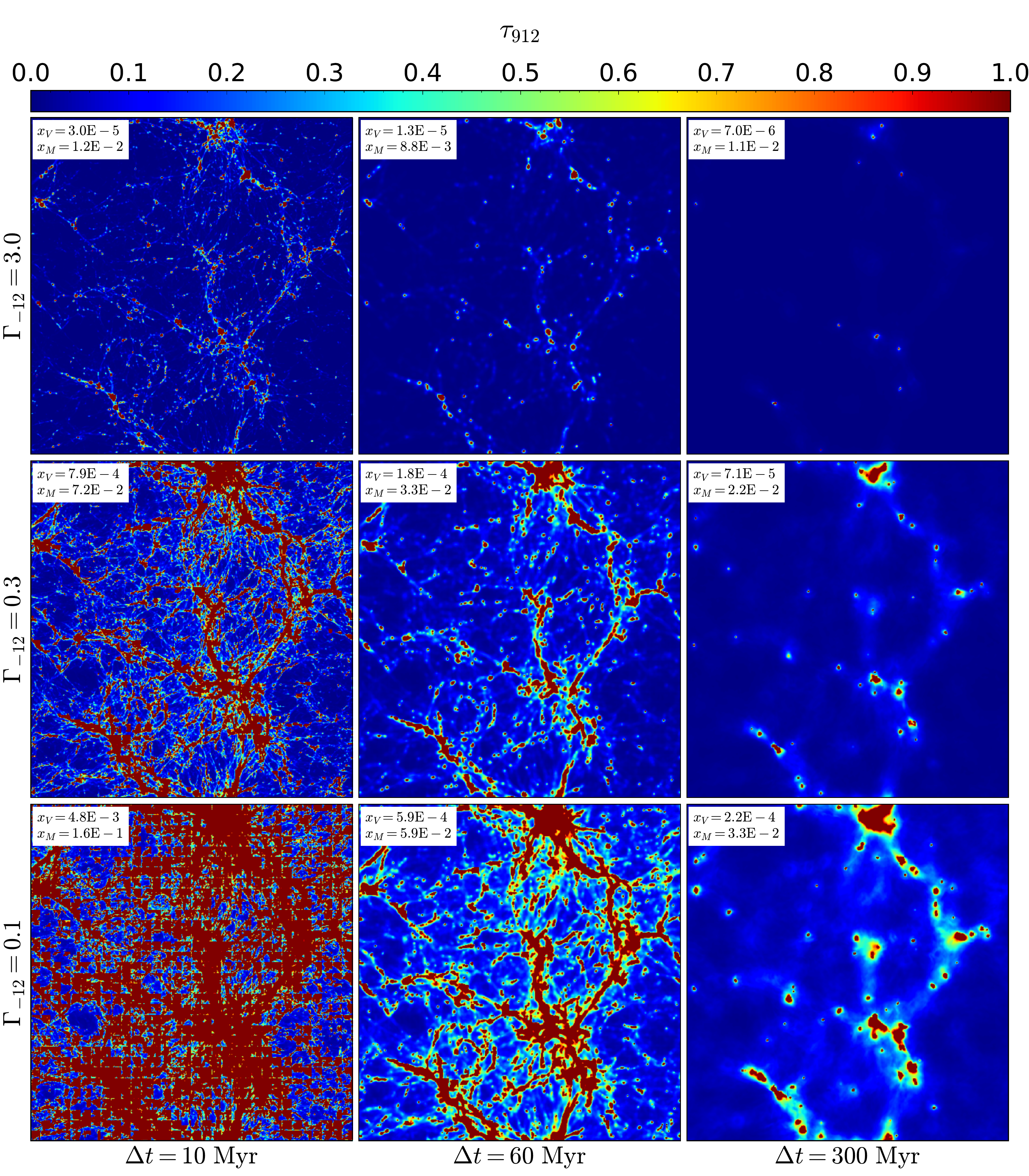} 
\caption{Evolution of self-shielding in the cosmic web after I-front passage.  Panels show $1\times 1$ $(h^{-1}\mathrm{Mpc})^2$ maps of the Lyman-limit ($\lambda = 912$\AA) optical depth integrated along the coordinate axis into the plane, with our color scheme saturating at $\tau_{912}=1$ to highlight self shielding regions.  The top, middle, and bottom rows correspond to $\Gamma_{-12}=3.0$, 0.3, and 0.1, respectively.   From left to right the columns correspond to $\Delta t = 10$, 60, and 300 Myr from $\zreion =8$. Each panel is annotated in the top-left corner with the corresponding volume- and mass-weighted average neutral fractions of the simulation volume.    At $\Delta t = 10$ Myr, the gas has not had time to relax, and a web of kiloparsec-scale structures shape the optical depth.  As time proceeds the self-shielding regions become smaller and disconnected.  The $\tau_{912} \geq 1$ regions are larger and more connected for lower $\Gamma_{-12}$. The grid-like structure most apparent in the $\Delta t = 10$ Myr, $\Gamma_{-12}=0.1$ panel is an artifact of our RT domain setup that the appendix shows does not affect our main results.}
\label{fig:tau912_visA}
\end{figure*}

\begin{figure*}
\includegraphics[width=18cm]{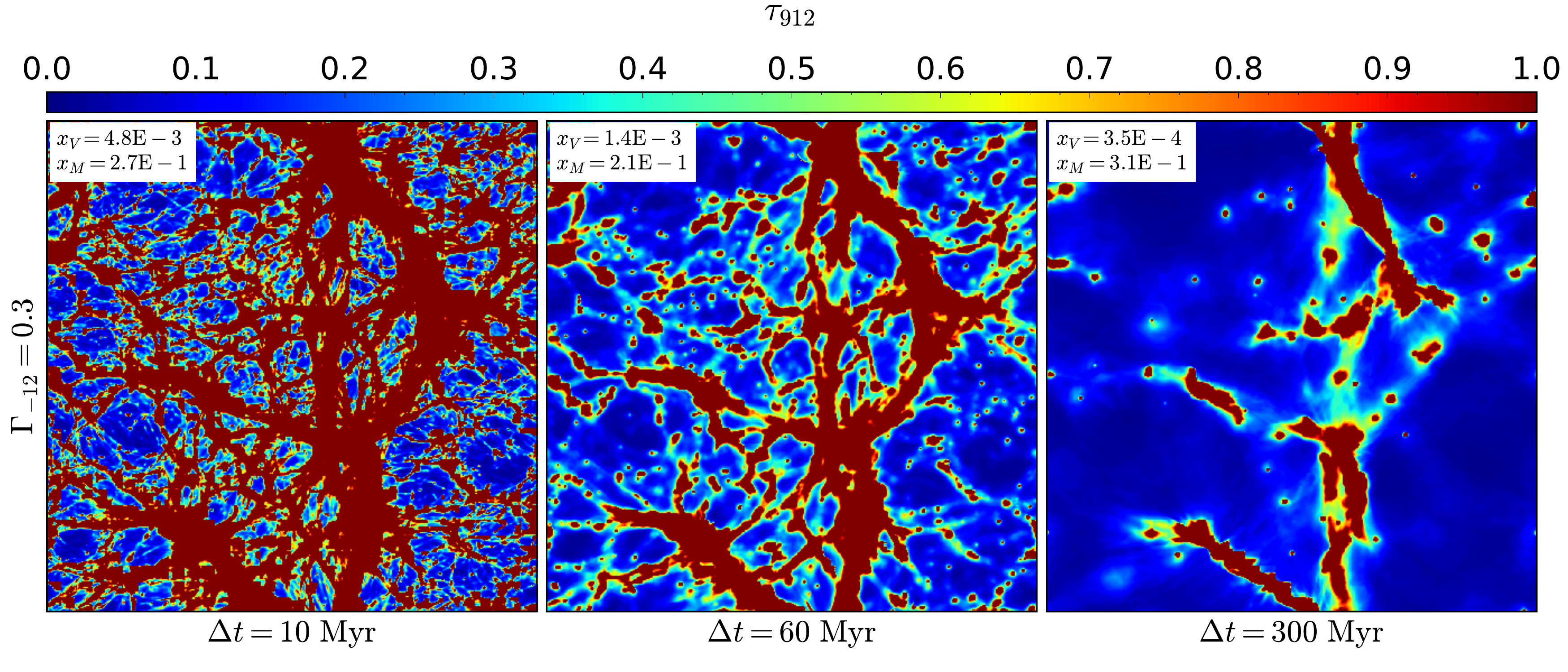} 
\caption{Same as in the middle panels of Fig. \ref{fig:tau912_visA} (with $\Gamma_{-12}=0.3$), but for the corresponding DC mode simulation with $\delta/\sigma = \sqrt{3}$.  Comparing to Fig. \ref{fig:tau912_visA} illustrates the impact of environmental density on the topology of self-shielding. The comoving side-length, from left to right, is $L_{\rm box} = 0.75$, $0.73$, and $0.65$ $h^{-1}$Mpc.  We have used the same random seed for all simulations.     }
\label{fig:tau912_visB}
\end{figure*}

 \subsection{Visualization of the density field}

 We begin with visualizations of the IGM during the relaxation process. Figure \ref{fig:gas_density} shows slices through the gas density in our simulation with $(\Gamma_{-12},\zreion,\delta/\sigma) = (0.3,8,0)$.  From top-left to bottom-right (going from left to right), the panels correspond to time intervals of $\Delta t = 10$, 60, 150 and 300 Myr from $\zreion = 8$.\footnote{In terms of redshift, these correspond to $z= 7.9$, $7.5$, $7.15$, and $6.0$.}  At $\Delta t = 10$ Myr since I-fronts swept through the box, the IGM has just begun to respond and the small-scale structure that formed in the unheated, pre-reionization gas remains intact.  By $\Delta t = 300$ Myr, the smallest structures have been erased; the gas has largely reached its relaxed state.  The panels exhibit weak shock fronts and conspicuous interference patterns formed by intergalactic sound waves. In each panel there is a characteristic scale corresponding to the ``sound horizon" $c_s \Delta t$, or the distance that a sound wave could have traveled since $\zreion$.

 \subsection{Visualization of self-shielding regions}
 \label{sec:SSRs}

 \begin{figure*}
\includegraphics[width=18cm]{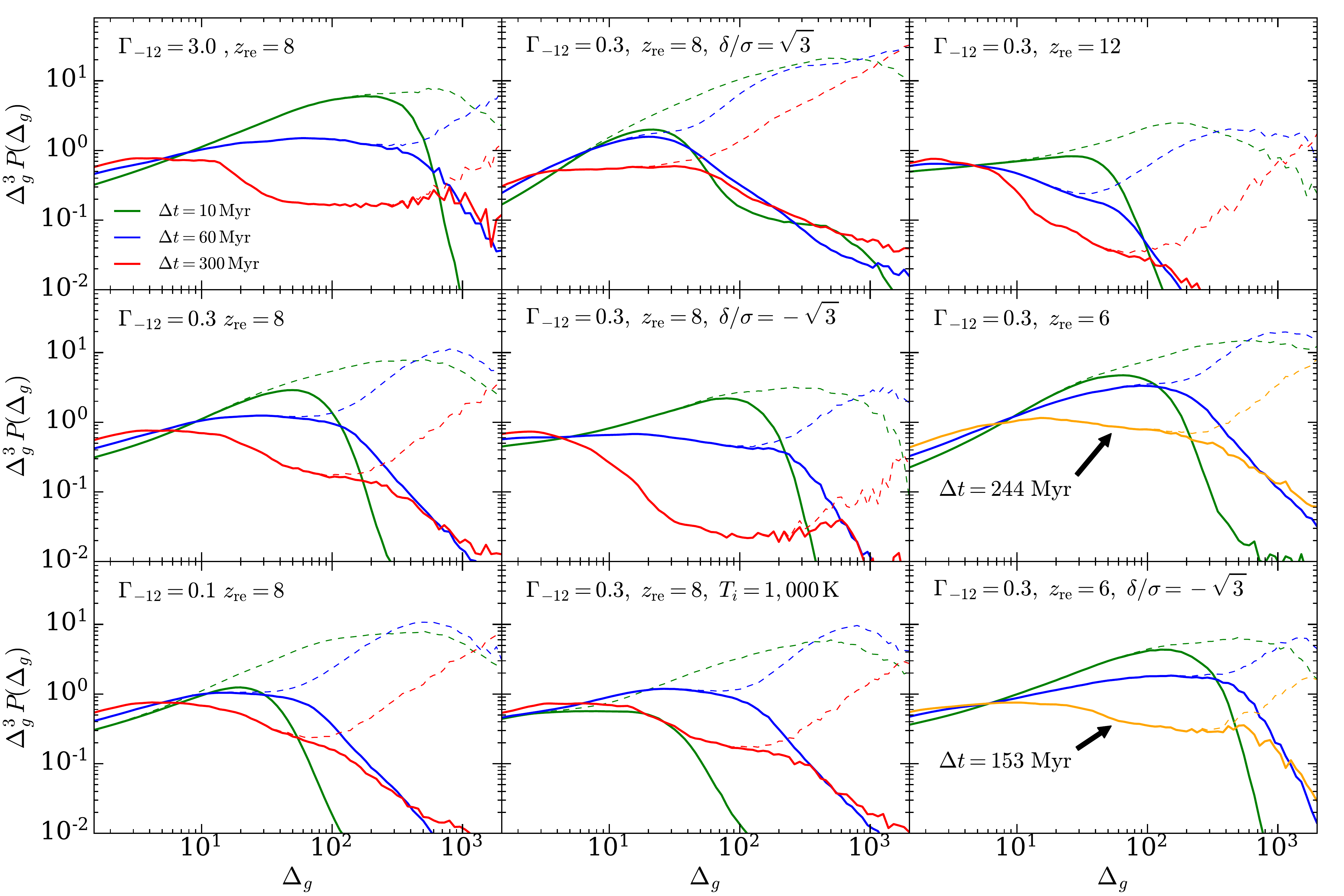} 
\caption{The distribution of gas densities in a selection of nine simulations with parameters denoted at the top of each panel. The green, blue, and red curves correspond to $\Delta t = 10$, 60, and 300 Myr from $\zreion$.  { For the two runs with $\zreion=6$ (middle- and bottom-right), the orange curves show results from the last simulation snapshot, as indicated by the labels.} The dashed curves show $\Delta_g^3 P(\Delta_g)$ for all gas, where $\Delta_g$ is the density in units of the mean.  The solid curves correspond to only ionized gas or, more precisely, $\bar{x}_i^2(\Delta_g) \Delta_g^3 P(\Delta_g)$, which is proportional to the contribution per logarithmic interval to the recombination rate (see main text).  The high-density cutoffs of the solid curves show the densities above which self-shielding occurs.  Self-shielding and relaxation significantly alter the density structure of the gas.    
\label{fig:ionPDF}}
\end{figure*}

\begin{figure}
\includegraphics[width=8.5cm]{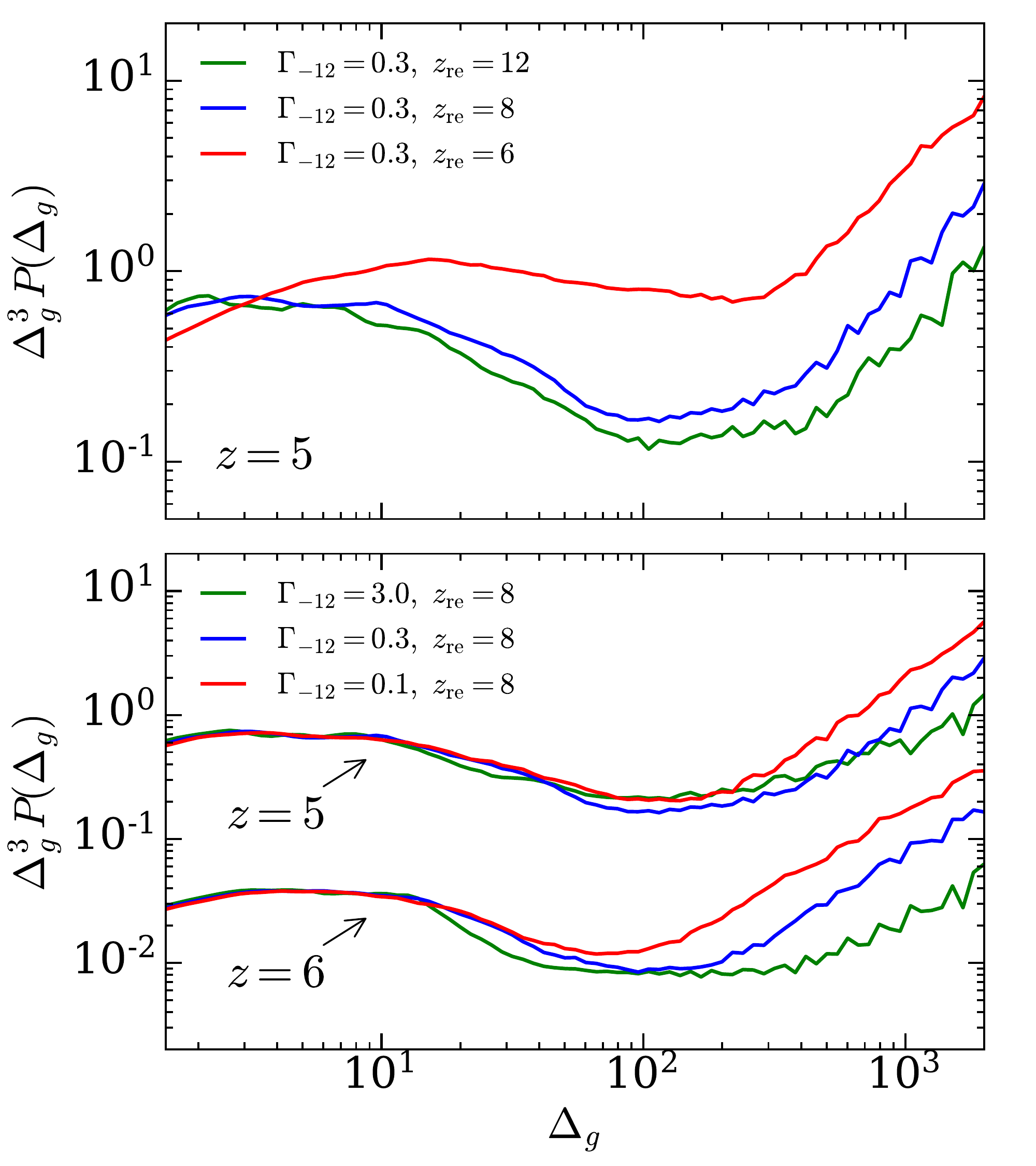} 
\caption{Memory of the local history of the ionizing background in the relaxed gas distribution.  Here we show $\Delta_g^3 P(\Delta_g)$ for all gas at fixed redshifts for $\Delta t \geq 300$ Myr after $\zreion$. The top panel explores the effect of the local timing of reionization by showing different $\zreion$ for fixed $\Gamma_{-12}$.  The bottom panel explores the effect of background intensity with different $\Gamma_{-12}$ and fixed $\zreion=8$.  In the bottom panel we have shifted the $z=6$ curves downward for clarity. Even in the relaxed limit well after ionization, the distribution of gas densities depends on the local history of the ionizing background.} 
\label{fig:ionPDFB}
\end{figure}
 
 Next we visualize the Lyman limit opacities along sight lines through our simulation boxes.  These visualizations show the structures within $\HII$ regions that are likely to absorb ionizing photons.  We construct two-dimensional maps of the opacity by computing $\tau_{912}$ for skewers of length $L_{\mathrm{box}}$, traced along the coordinate axis into the page.   Figure \ref{fig:tau912_visA} illustrates how the maps evolve in time. The colors represent $\tau_{912}$ and we have set an upper limit of $\tau_{912}=1$ in order to highlight the self shielding structures.  The top, middle, and bottom rows correspond to $\Gamma_{-12} = 3.0$, $0.3$, and $0.1$, while the left, middle, and right columns show $\Delta t = 10$, 60, and 300 Myr, respectively.  In all cases, the ionizing radiation turns on at $\zreion = 8$.  The top-left corner of each panel displays the corresponding volume- and mass-weighted mean neutral fraction of the entire simulation volume, denoted $x_V$ and $x_M$, respectively.     Additionally, Fig. \ref{fig:tau912_visB} shows a $\tau_{912}$ map from one of our DC mode runs with $\delta/\sigma = \sqrt{3}$.  Note that our use of the same initial seeds for all runs allows a direct comparison of structures.    
 
 Consider the middle row of Fig. \ref{fig:tau912_visA}.  At $\Delta t = 10$ Myr, the map exhibits small-scale structure and the covering fraction of $\tau_{912} \ge 1$ is large.  The $\tau_{912}$ regions extend well beyond halos; indeed, they are connected by filaments.  The small-scale structure is erased as self-shielding clumps are photoevaporated.  By $\Delta t = 300$ Myr, the remaining $\tau_{912} \ge 1$ regions are relegated to rare, high-density peaks around massive halos (right panel).     
 
 Comparing the rows of Fig. \ref{fig:tau912_visA} shows visually how the morphology of self-shielding depends on the intensity of the ionizing radiation background.  For $\Gamma_{-12}=3.0$, the $\tau_{912}\ge 1$ regions are already isolated to small islands by $\Delta t = 10$ Myr, whereas even $\Delta_g = 10$ gas is able to self-shield at this time for $\Gamma_{-12}=0.1$.  The grid-like structure seen for the latter (and more subtly in the other $\Delta t = 10$ Myr panels) is a relic of our RT domain setup (see \S \ref{sec:methods}).  We emphasize that the box is already highly ionized by $\Delta t = 10$ Myr for all cases shown here.

 Figure \ref{fig:tau912_visB} shows how the self-shielding changes with environmental density. The over-dense ($\delta/\sigma=\sqrt{3}$) run is effectively a more evolved version of the corresponding $\delta/\sigma=0$ run shown in the middle panels of Fig. \ref{fig:tau912_visA}. The structure is more non-linear in the former and the consequence of this can be seen in the increased amount of self-shielding.

 \subsection{Self-shielding and the distribution of gas densities}
 \label{sec:ionPDF}
 
 In Fig. \ref{fig:ionPDF} we examine the self-shielding more quantitatively with the probability distribution of ionized gas density.  The dashed curves show $\Delta_g^3 P(\Delta_g)$, where $P(\Delta_g)$ is the volume-weighted probability distribution of $\Delta_g$.  Following \citet{2011ApJ...743...82M}, the solid curves show this quantity multiplied by the square of the mean ionized fraction at density $\Delta_g$, $\bar{x}_{\rm HII}^2(\Delta_g)$.  Note that $\bar{x}^2_{\rm HII}(\Delta_g) \Delta_g^3 P(\Delta_g)$ is proportional to the contribution of gas at density $\Delta_g$ to the global recombination rate, per logarithmic interval in $\Delta_g$.  In other words, the recombination rate can be written as $R \propto \int_0^{\infty} d\ln \Delta_g \bar{x}_{\rm HII}^2(\Delta_g) \Delta_g^2 P(\Delta_g)$.  The often sharp transitions to zero exhibited by the solid curves at high $\Delta_g$ owe to self-shielding. 
 
Figure \ref{fig:ionPDF} explores different values of $\Gamma_{-12}$, $\zreion$, and $\delta/\sigma$, as marked in each panel ($\delta/\sigma=0$ unless otherwise stated).  In the center-bottom panel we show results from our pre-heating run with $T_i=1,000$ K.  The green, blue, and red curves correspond to cosmic time intervals of $\Delta t = 10$, 60, and 300 Myr measured from $z=\zreion$.  { For the two runs with $\zreion=6$ (middle- and bottom-right), the orange curves show results from the last simulation snapshot, as indicated by the labels}.  

Clearly the gas distribution and the self-shielding densities depend on the intensity of the ionizing background, the timing of (local) reionization, and the environmental density.   But the evolution is qualitatively similar in all cases. The IGM contains more gas at $\Delta_g > 10$ before relaxation. The I-fronts penetrate into density peaks, with a depth determined by $\Gamma_{-12}$, and the recombination rate reaches a maximum.  By $\Delta t = 300$ Myr, much of the gas has been evacuated to lower density such that the (now lower) recombination rate receives a larger fractional contribution from $\Delta_g = 1-10$.

Does the post-reionization gas distribution depend on the details of the local reionization history? We address this question in Fig. \ref{fig:ionPDFB} by showing $\Delta_g^3 P(\Delta_g)$ at fixed redshifts for different $\Gamma_{-12}$ (bottom) and $\zreion$ (top).  The bottom panel shows that, for fixed $\zreion = 8$, a more intense local background leads to less gas at high density.  The top and bottom sets of curves in that panel correspond to snapshots at $z=5$ and $z=6$, respectively, where we have re-scaled the latter by a factor of $1/20$ for clarity.  The bottom panel further illustrates that gravitational collapse eventually drives the distributions to a more similar shape as the high-density regime becomes dominated by more-massive halos.  The  differences among the curves are more pronounced at $z=6$ than at $z=5$.

Considering now the impact of $\zreion$ at fixed $\Gamma_{-12}$, the top panel of Fig. \ref{fig:ionPDFB} shows that regions with lower $\zreion$ retain more high-density gas.  This effect, which is particularly evident in the $\zreion=6$ case, can be attributed to two causes: (1) The gas spent more of its time evolving at cold, pre-reionization temperatures, during which structure formation was unimpeded by pressure smoothing. More gas was able to collapse to high, self-shielding densities.  This effect is responsible for the differences between the $\zreion = 8$ and 12 cases, which are both in the relaxed limit by $z=5$; (2) For the $\zreion=6$ case, the dominant effect is that the gas is also still undergoing relaxation at $z=5$ ($\Delta t = 240$ Myr; we will show this in the next section).  

A central finding of this work is that the distribution of gas density can vary with location due to the local history of the ionizing background. Imagine two similar patches of the IGM exposed to radiation from two different star-formation histories, e.g. nearby starbursts at different cosmic times and/or intensities.  Even well after reionization, differences in the two histories can manifest themselves in the gas distributions because of the interplay between self-shielding and the relaxation process. Gravity acts to erase much of this variation as the high-density regime becomes dominated by larger shelf-shielding structures over a Hubble time.

\subsection{The Recombination Rate}
\label{sec:recrate}

  We quantify the recombination rate in our simulation boxes with a clumping factor,

\begin{equation}
 C_\mathrm{R} =  \frac{\langle  \alpha_\mathrm{B}(T) n_\HII  n_\mathrm{e} \rangle}{ \alpha_\mathrm{B}(T_{\mathrm{ref}}) \langle n_\HII \rangle \langle n_\mathrm{e} \rangle},
 \label{eq:clumping}
 \end{equation}
 where $T_{\mathrm{ref}}$ is a reference temperature that we take to be $10,000$ K, and the angular brackets denote spatial averages over the simulation box. Physically, $C_R$ is the factor by which the recombination rate is boosted by the presence of small-scale structure, over the (hypothetical) case of a homogeneous density field with uniform temperature $T_{\mathrm{ref}}$.

  \begin{figure}
\includegraphics[width=8.5cm]{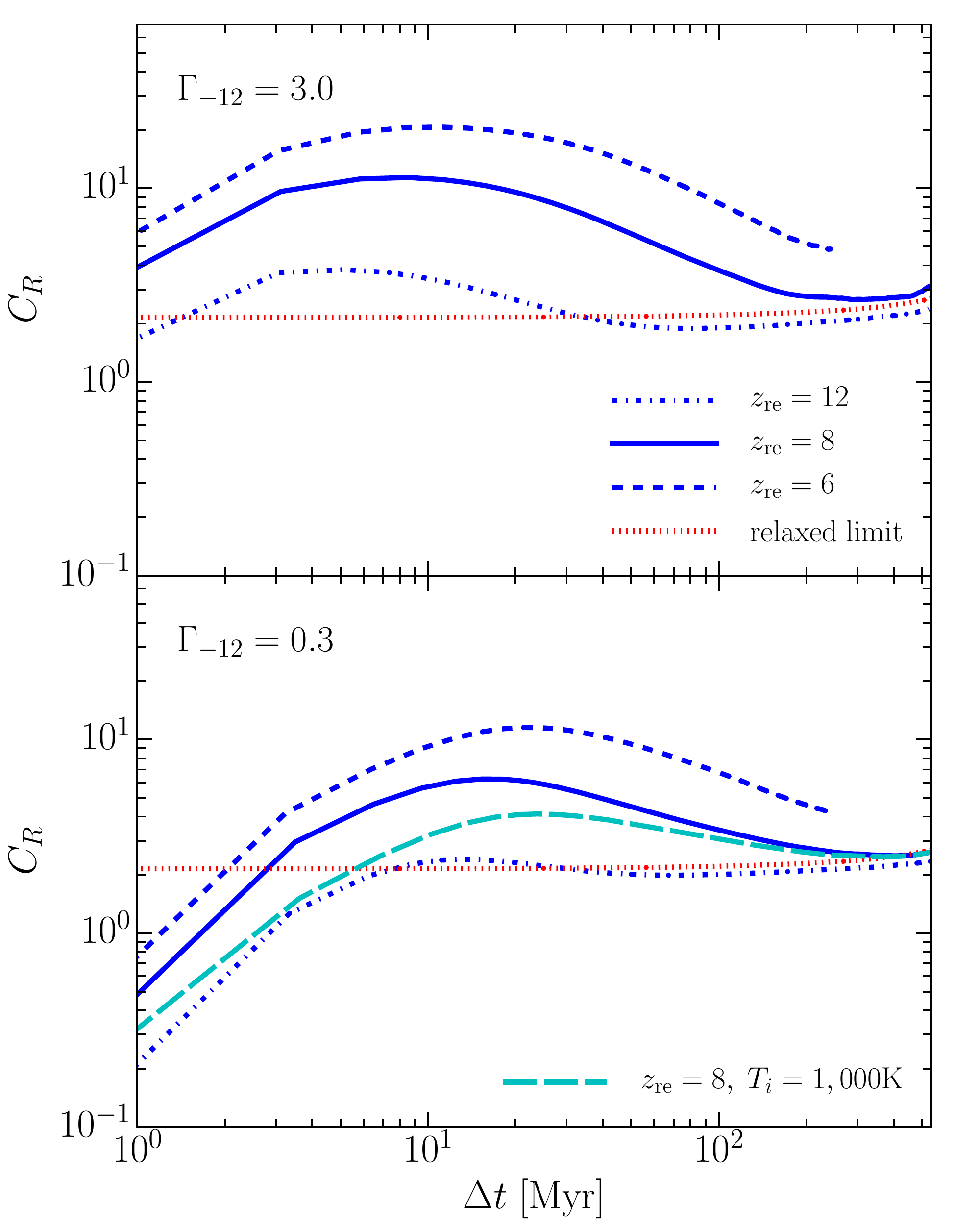} 
\caption{Evolution of the clumping factor as the gas relaxes in response to reionization. The top and bottom panels correspond to $\Gamma_{-12} = 3$ and 0.3, respectively.  The blue curves correspond to different $\zreion$ as denoted in the top panel. The red/dotted curves show the relaxed (or late-time) limit of the clumping factor, which is nearly independent of $z$, $\Gamma_{-12}$ and $\zreion$ (see text for more details).  In the bottom panel, the cyan/long-dashed curve shows results from our pre-heating run with a $1,000$K temperature floor at $z<20$, which we argue in the main text should provide an upper bound on the effect of preheating by early X-ray sources. }
\label{fig:clumping}
\end{figure}

Figure \ref{fig:clumping} shows how the clumping factor evolves during the relaxation process. In the bottom (top) panel we show $C_R$ as a function of cosmic time since $\zreion$ for $\Gamma_{-12} = 0.3$ (3.0).  The blue dot-dashed, solid, and short-dashed curves correspond to $\zreion = 12$, 8, and 6, respectively. (The $\zreion=6$ curves terminate earlier because we did not run any of these simulations past $z=5$.)  The cyan/long-dashed curve in the bottom panel corresponds to our $T_i=1,000$ K run, which serves as a crude upper limit on the impact of pre-heating by the first X-ray sources.  

As the I-fronts sweep through the domains, the ionizing radiation races up the over-densities ahead of their hydrodynamic response, causing the recombination rate to peak at $\Delta t \approx 10$ Myr (or somewhat earlier in the $\Gamma_{-12}=3.0$ case because of the faster I-fronts). Larger $\Gamma_{-12}$ produces a higher peak $C_R$ because the radiation is able to penetrate more deeply into the over-densities. Lower $\zreion$ produces a higher peak because structure formation has had more time to amplify the over-densities before they are photoheated.  The clumping factor declines after $\Delta t \approx 10$ Myr as over-dense gas is photoevaporated to lower densities.

 It is useful to compare $C_R(t)$ to its late-time evolution, i.e. after the gas has adjusted to the heating from reionization. To this end, we obtain ``relaxed limit" models, shown as the red/dotted curves, using our $\zreion=12$ runs.  For these curves, $\Delta t$ measures the cosmic time since $z=8$.  However, we find that $C_R$ in this late-time regime is very weakly dependent on time\footnote{For over-dense boxes, however, the time-dependence is stronger} and $\Gamma_{-12}$, so the red/dotted curves serve as a general basis of reference.    
 
 Our results indicate that $C_R$ eventually approaches approximately the same relaxed limit, which we find is remarkably insensitive to $\Gamma_{-12}$. At first glance this result may be surprising because the densities that self-shield are very different between $\Gamma_{-12}=0.3$ and $\Gamma_{-12}=3.0$, for example (see the $\Delta t = 300$ Myr curves in the top- and middle-left panels of Fig. \ref{fig:ionPDF}).  The insensitivity to $\Gamma_{-12}$ results from two effects; (1) At late times, densities of $\Delta_g = 1-10$ contribute relatively more to the recombination rate, and the gas distributions at these densities are nearly identical.  This can be best seen in the bottom panel of Fig. \ref{fig:ionPDFB}; (2) In the case of higher $\Gamma_{-12}$, photoevaporation removes more gas from moderate and high densities.  At late-times, the lack of gas at these densities compensates for the higher threshold for self-shielding.

 A comparison of the cyan/long-dashed and blue/solid curves in the bottom panel of Fig. \ref{fig:clumping} reveals that the impact of pre-heating is greatest during the early phases of the relaxation process.  The pre-heating raises the $z>\zreion$ Jeans temperature, which erases some of the small-scale density peaks that play a role in the early evolution of $C_R$. As a result, $C_R$ is $\lesssim 80$ \% lower within the first 10 Myrs.  Note also that the peak in $C_R$ occurs $\approx 10$ Myr later compared to the fiducial run (solid/blue) because the photoevaporaton time-scale for the more massive clumps that survive pre-heating is longer \citep{2004MNRAS.348..753S, 2005MNRAS.361..405I}. 
 
 The trends found here are qualitatively similar to those reported in \citet{2016ApJ...831...86P}, though our results differ in detail. The most direct comparison can be drawn with their M\_I-0.5\_z10 run, for which $\Gamma_{-12} = 2.8$ and $\zreion=10$. Examining the solid and dashed curves in the top panel of Fig. \ref{fig:clumping}, our $C_R$ reaches a similar maximum value of $10-15$ (compared to 13 in their simulation).  However, the peak is reached at $\Delta t \approx 8$ Myr, which is significantly later than the $\Delta t \approx 3$ Myr peak found in \citet{2016ApJ...831...86P}. In addition,  our $C_R$ relaxes to a value of $\approx 3$, whereas \citet{2016ApJ...831...86P} found $C_R\approx 1.5$ at $\Delta t = 150$ Myr.\footnote{We note that this comparison is not exact, as \citet{2016ApJ...831...86P} take $T_{\rm ref}$ to be the mass-weighted average of their simulation.    } These differences may owe to their approximate model for RT.

 \subsection{Mean Free Path}
 
 \begin{figure}
\includegraphics[width=8.5cm]{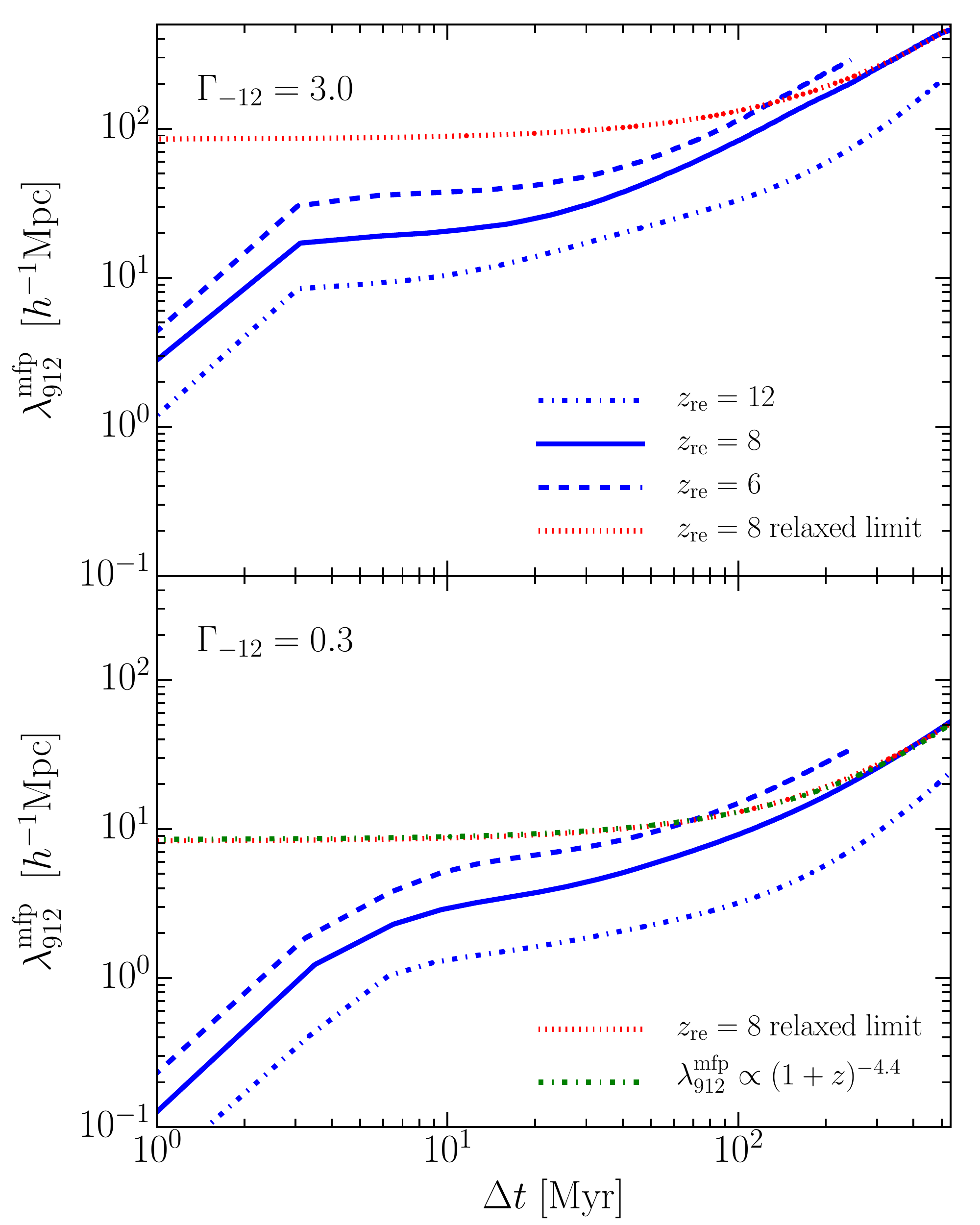} 
\caption{Evolution of the (local) mean free path during relaxation.  Line styles are the same as in Fig. \ref{fig:clumping}.  In contrast to the clumping factor, the relaxed limit evolves with $z$ and is nearly proportional to $\Gamma_{-12}$.  Thus, for clarity, we show the relaxed limits for the $\zreion=8$ cases only.  The time-evolution of the relaxed limit is well-described by the empirical scaling $\MFP \propto (1+z)^{-4.4}$ of \citet{2014MNRAS.445.1745W}, which we show in the bottom panel as the green/dot-dashed curve.   }
\label{fig:MFP}
\end{figure}

 We computed the mean free path using the method of \citet{2013ApJ...763..146E}.  To maximize the use of the simulation volume, we filled the box with skewers traced along the three coordinate axes.  The skewers were divided into segments of length $L_{\mathrm{dom}}=32~h^{-1}$ kpc.  For each segment we measured the fraction of transmitted flux at wavelength 912 \AA, $f_{\mathrm{out}} = \exp(-\tau_{912})$, where $\tau_{912}=\int_0^{L_{\mathrm{dom}}} ds~n_{\mathrm{HI}}(s)\sigma_{912}$, and $\sigma_{912}$ is the hydrogen photoionization cross section evaluated at 912 \AA. We then averaged  $f_{\mathrm{out}}$ over all segments and computed the mean free path using $\lambda^{\mathrm{mfp}}_{912} = -L_{\mathrm{dom}}/\ln( \langle f_{\mathrm{out}}\rangle)$.  We note that this procedure breaks down for $\lambda^{\mathrm{mfp}}_{912}\lesssim L_{\rm dom}$, when skewer segments become optically thick. Fortunately, we will see that $\lambda^{\mathrm{mfp}}_{912}\gg L_{\rm dom}$ over almost all of the parameter space of interest. We have also checked this procedure for measuring $\MFP$ against an alternative method based on averaging $n_{\rm HI}\Gamma$.  We found good agreement between the two methods. 
 
 Figure \ref{fig:MFP} shows how $\MFP$ evolves during the relaxation process.   The red/dotted curves correspond to relaxed limits.  In contrast to the clumping factor, we find that the relaxed limit of $\MFP$ evolves quickly with time.   Hence, for clarity, we show the relaxed limits for only the $\zreion=8$ runs.  Consistent with our discussion of self-shielding regions in \S $\ref{sec:SSRs}$, $\MFP$ starts out at its shortest during the un-relaxed phase following the passage of the I-fronts.  The mean free path rises as the gas responds, reaching the relaxed limit on a time scale of $\Delta t \approx 300$ Myr.  Fig. \ref{fig:MFP} also shows that $\MFP$ begins at larger values for lower $\zreion$.  This results from the combined effects of a decreasing mean density and structure formation moving gas into high-density peaks.  At later times self-shielding systems occupy a smaller fraction of the volume.  
 
 Whereas the relaxed limit of $C_R$ is characterized by a nearly constant value of $2-3$ (Fig. \ref{fig:clumping}), we find that the time-evolution of the relaxed limit is reproduced well by a power-law of the form $\MFP\propto (1+z)^{-4.4}$, which we obtained by fitting to the $\zreion = 12$ simulation results.  For comparison, we show this power-law scaling as the green/dot-dashed curve in the bottom panel of Fig. \ref{fig:MFP}.  Note that  $\MFP\propto (1+z)^{-4.4}$ (where the $\MFP$ here is comoving) is also the best-fit redshift dependence from \citet{2014MNRAS.445.1745W} obtained from observational measurements at $z \leq 5.2$.   This empirical evolution is well reproduced by the relaxed limit of our ``minimalist" simulations in which only hydrodynamics and RT shape the mean free path. 
 
 We now turn to how $\MFP$ depends on $\Gamma_{-12}$.  The steepness of this dependence determines how rapidly the opacity of the forest evolves with the emissivity of ionizing sources \citep{2011ApJ...743...82M}. Furthermore, a strong dependence of the {\it local} $\MFP$ on $\Gamma_{-12}$ can yield large fluctuations in the ionizing radiation background. These effects have been invoked in models of the ionizing background to explain the rapid evolution and large opacity fluctuations in the $z>5.5$ Ly$\alpha$ forest \citep{2015arXiv150907131D,2018MNRAS.473..560D,2019arXiv191003570N}.   
 
  A comparison between the red/dotted curves in the top and bottom panels of Fig. $\ref{fig:MFP}$ indicates that the relaxed $\MFP$ is, to a good approximation, proportional to $\Gamma_{-12}$. At face value, this result appears to imply a steeper scaling than the $\MFP \propto \Gamma_{-12}^{0.66-0.75}$ dependence found by \citet{2011ApJ...743...82M}.  However, the interpretation of this comparison is complicated by the fact that our mean free paths are calculated from simulations that have a single value of $\Gamma_{-12}$ over their entire ionization histories, whereas the scaling of \citet{2011ApJ...743...82M} was obtained in the limit of instantaneous change in $\Gamma_{-12}$ for fully relaxed gas.  The latter limit is more appropriate for the ionizing background model of \citet{2015arXiv150907131D}.

 We can nonetheless attempt to make more direct contact with the results of \citet{2011ApJ...743...82M} by examining the slopes of the relaxed gas density PDFs in Fig. \ref{fig:ionPDF}, i.e. the curves with $\Delta t = 300$ Myr.  The slopes allow one to estimate the scaling from an \emph{instantaneous} change in $\Gamma_{-12}$.  For all cases, we find that $P(\Delta)\sim \Delta^{-\gamma}$, with $\gamma \approx 1.7-1.8$, provides a reasonable approximation to the PDFs above the densities where self-shielding kicks in (which are the densities applicable for calculating the response to an increase in the ionizing background).  According to the arguments of \citet{2011ApJ...743...82M}, this implies a scaling of $\MFP \propto \Gamma_{-12}^{0.33}$, which is notably shallower than what they found. A similar scaling appears to hold during the relaxation process.  This shallower scaling would reduce fluctuations in the $z>5$ ionizing background relative to current models \citep{2015arXiv150907131D, 2019arXiv191003570N}. We caution, however, that the scaling is likely to be steeper than we estimate here because $\gamma$ becomes larger (such that $\Delta_g^3 P(\Delta_g)$ flattens out) near the densities of the self-shielding transition.  In future work, we plan to investigate the implications of these findings in more detail.

 \subsection{Effect of box-scale density fluctuations}
 \label{sec:DCmode}
 
 \begin{figure*}
 \begin{center}
\includegraphics[width=7.5cm]{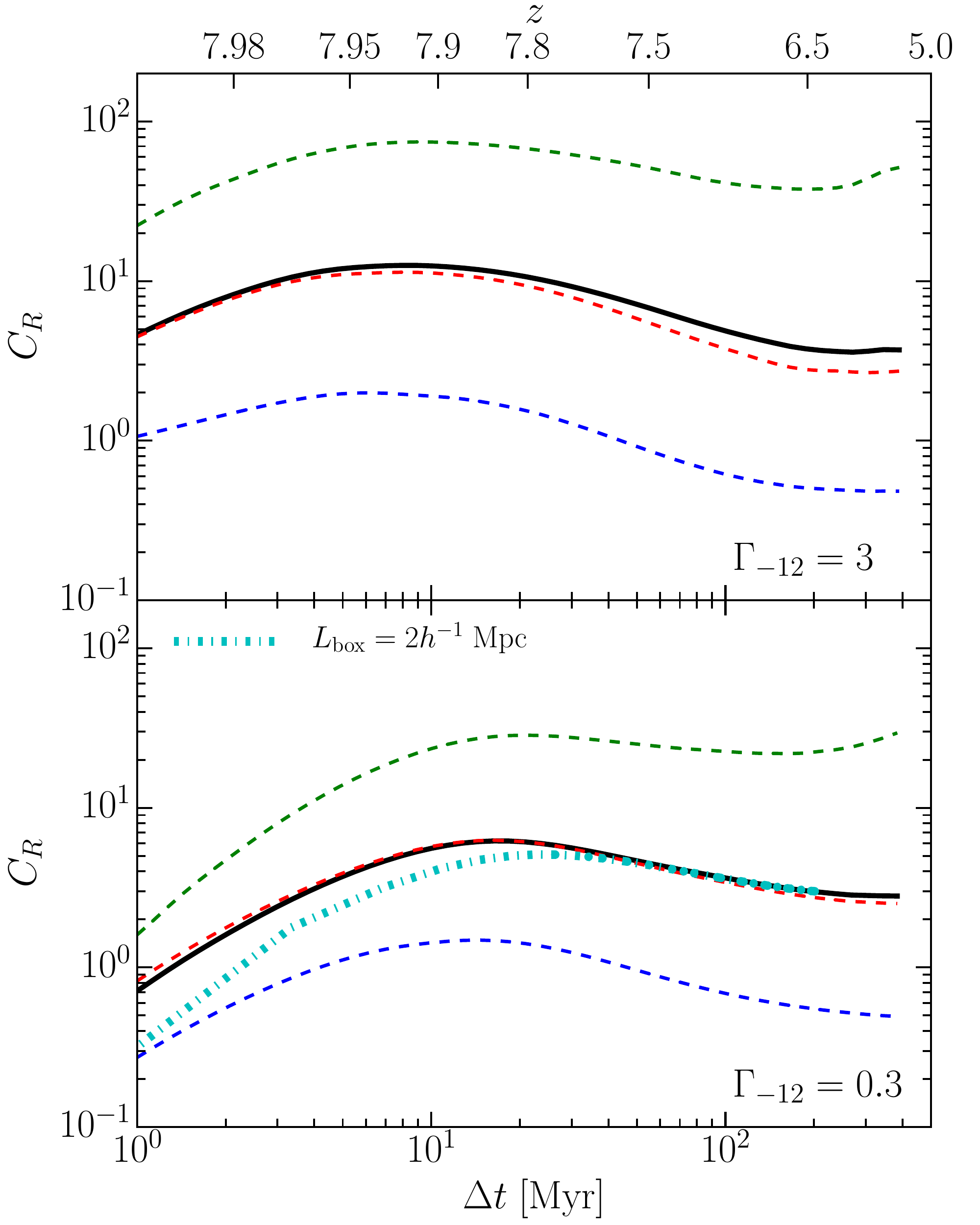}
\includegraphics[width=7.5cm]{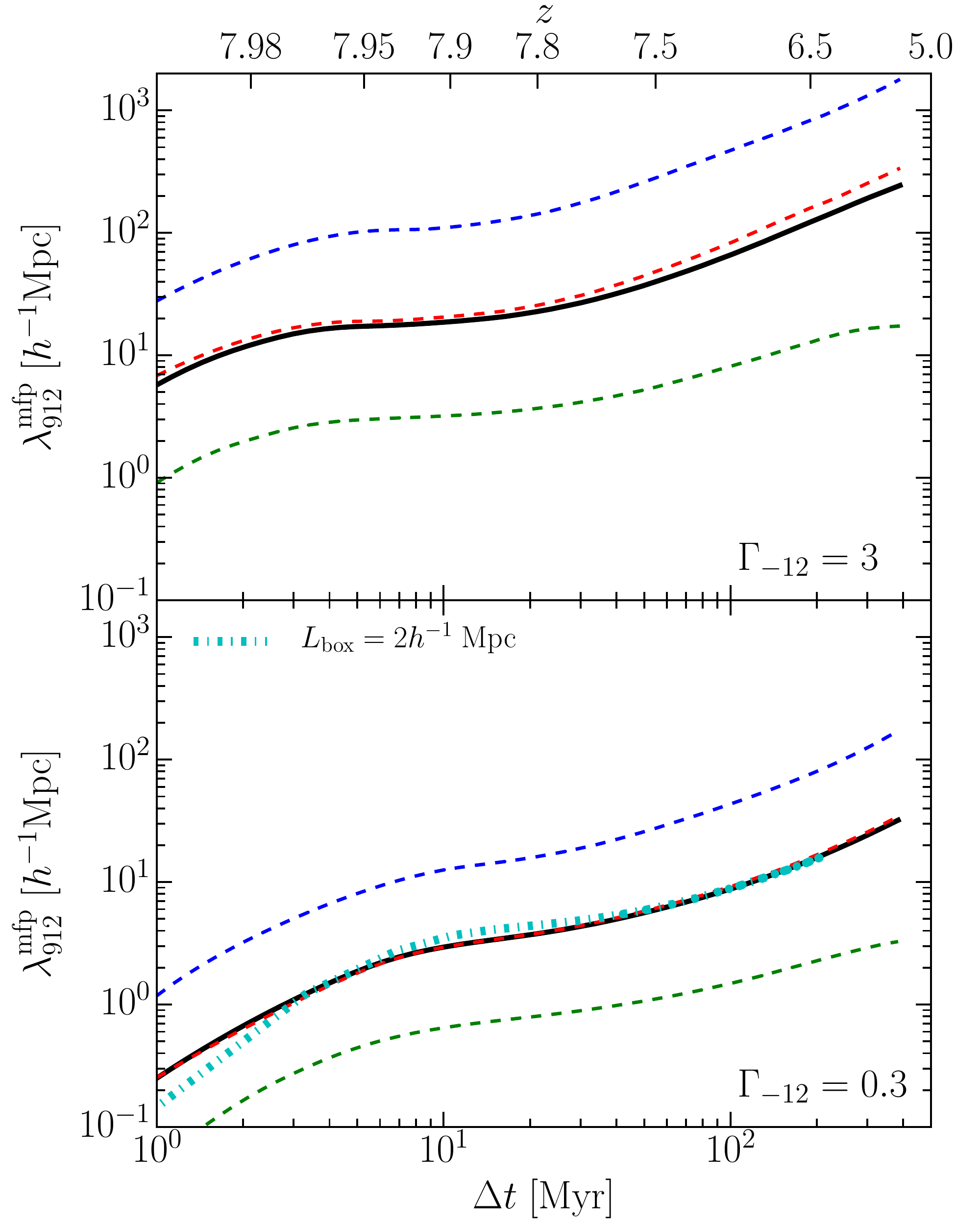}
\end{center}
\caption{The effect of box-scale density fluctuations on the clumping factor (left panels) and mean free path (right panels).  Here we show results for fixed $\zreion =8$.  The green, red, and blue dashed curves correspond to $\delta/\sigma = +\sqrt{3}$, 0, and $-\sqrt{3}$, respectively. The solid black curves show the average over all box-scale densities using our Gauss-Hermite Quatrature method.  The agreement between the average and our $\delta/\sigma = 0$ run indicates that the mean-density simulation is a good approximation for $\langle C_R \rangle$ and $\langle \MFP \rangle$.  To cross-check our averaging scheme, the bottom panels show results from a simulation with 2 $h^{-1}$Mpc, $N=1024^3$, which has 8 times the volume of our production runs (but with 1/8th the resolution elements).}
\label{fig:int_clumping}\label{fig:MFP_integrated}
\end{figure*}

 In Fig. \ref{fig:int_clumping}, we use our DC mode simulations to examine the impact of box-scale density variations on the local and $\delta$-averaged $C_R$ (left) and $\MFP$ (right).  The green, red, and blue dashed curves correspond to $\delta/\sigma = +\sqrt{3}$, 0, and $-\sqrt{3}$, respectively. { In the DC mode runs, $C_R$ quantifies the boost in the recombination rate above the cosmic mean.  In other words, we measure the numerator in equation (\ref{eq:clumping}) from the over/under-dense simulations, but take $\langle n_{\mathrm{HII}}\rangle$ and $\langle n_e \rangle$ in the denominator to be the cosmic mean values.} The solid curves correspond to the $\delta$-averaged quantities, $\langle C_R(t)\rangle$ and $\langle \MFP(t) \rangle$, calculated from Gauss-Hermite Quadrature using equation (\ref{eq:quadrature}).  All results are for $\zreion=8$ and the bottom and top panels illustrate the effect of varying $\Gamma_{-12}$ from 0.3 to 3.0.  

As noted above, the DC mode runs can be thought of as universes in which structure formation is more advanced ($\delta/\sigma>0$) or delayed ($\delta/\sigma<0$).  In the over-dense (under-dense) simulation, the larger (smaller) variance of density fluctuations results in a boosted (suppressed) recombination rate, as shown in the left panels of Fig. \ref{fig:int_clumping}. Likewise, the abundance (paucity) of self-shielding regions in the over-dense (under-dense) runs result in a shorter (longer) mean free path.

In \S \ref{sec:methods} we raised the concern that our $\delta/\sigma=0$ runs miss potentially important effects of large density fluctuations on our box scale.  What impact do these fluctuations have on the recombination rate and mean free path?   We find that both $\langle C_R(t) \rangle$ and $\langle \MFP(t) \rangle$ are quite close to the results from our $\delta/\sigma = 0$ runs. (Compare solid and the red dashed curves in each of the panels in Fig.~\ref{fig:int_clumping}.)   The similarity owes to a large cancellation between the contributions of over- and under-dense regions. We emphasize that, although we have utilized three simulations to evaluate the integral (\ref{eq:genGI}), our Gauss-Hermite Quadrature method approximates the average over the entire range of box-scale overdensities. It appears, then, that our $\delta/\sigma=0$ runs match reasonably well what we would have obtained if it were feasible to run a larger simulation.

One might still worry that our Gauss-Hermite Quadrature method misses the contribution of rare over-dense absorbers that are not captured in our simulation volumes.  Indeed, in the case of the mean free path, for example, our method relies on approximating the quantity $(\MFP)^{-1}/(1+\delta_{\rm NL})$, where $\delta_{\rm NL}$ is the non-linear DC mode of the box, as a 5th-order polynomial.\footnote{In fact, if this quantity were a 5th-order polynomial, then our computation of the integral, eq. (\ref{eq:genGI}), would be exact.} We have not observed any evidence that this approximation fails.  To support this statement, the cyan/dot-dashed curve in the bottom panels of Fig. \ref{fig:int_clumping} show $C_R$ (left) and $\lambda^{\rm mfp}_{912}$ (right) from a run with $N = 1024^3$ and $L_{\rm box}=2.048 h^{-1}$ Mpc, eight times the volume of our fiducial runs.  This run has poorer resolution, so we expect some deviation early in the evolution owing to numerical convergence (see discussion in Appendix \ref{sec:convergence}).  However, the agreement, especially at $\Delta t \gtrsim 20$ Myr, is encouraging indication that our averaging method picks up the contribution from structures missed by our $L_{\rm} = 1.024 h^{-1}$ Mpc ($\delta/\sigma=0$) boxes.

 \section{Implications for Reionization}
 \label{sec:implications}

\begin{figure*}
\includegraphics[width=8.5cm]{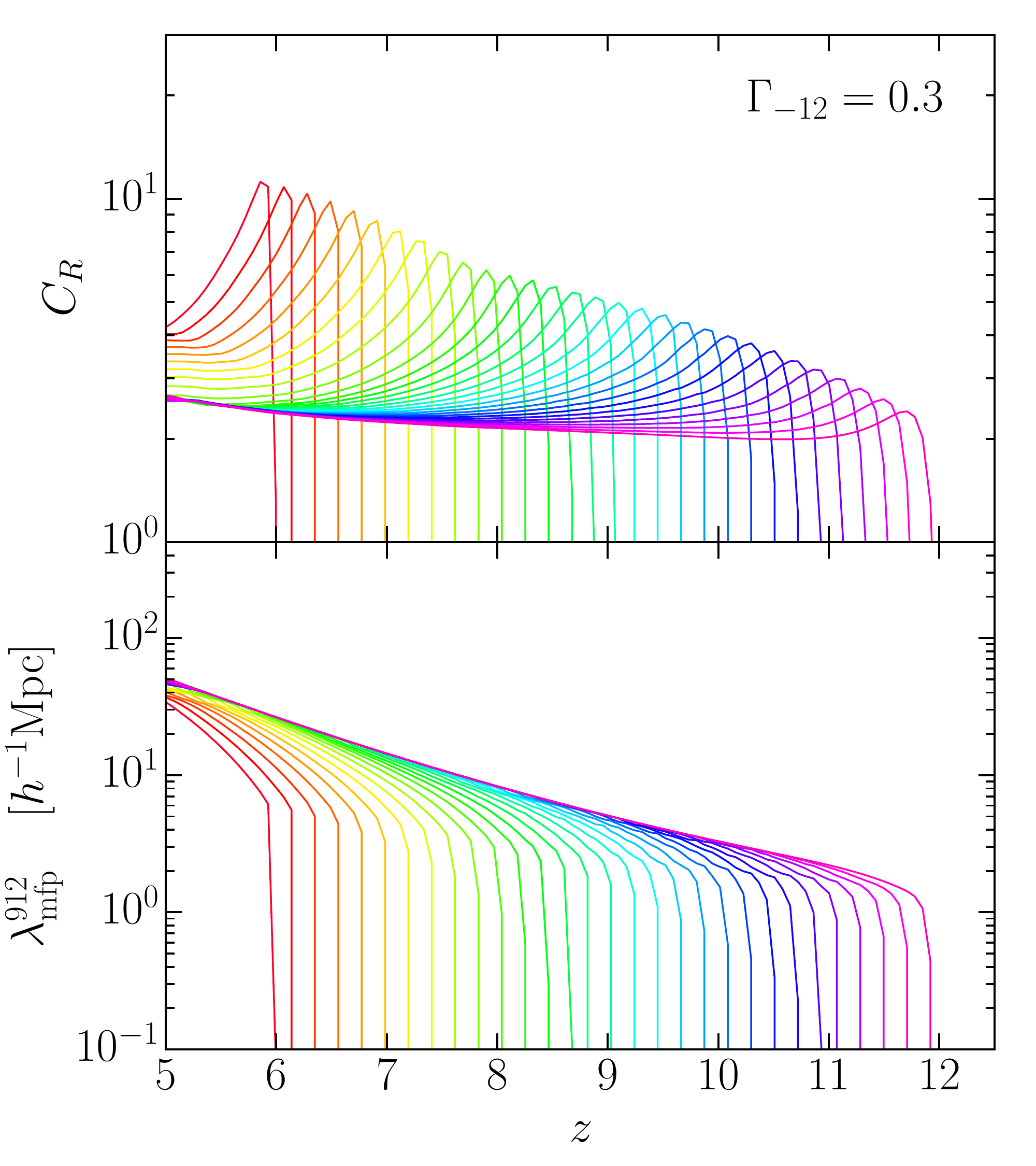} 
\includegraphics[width=8.5cm]{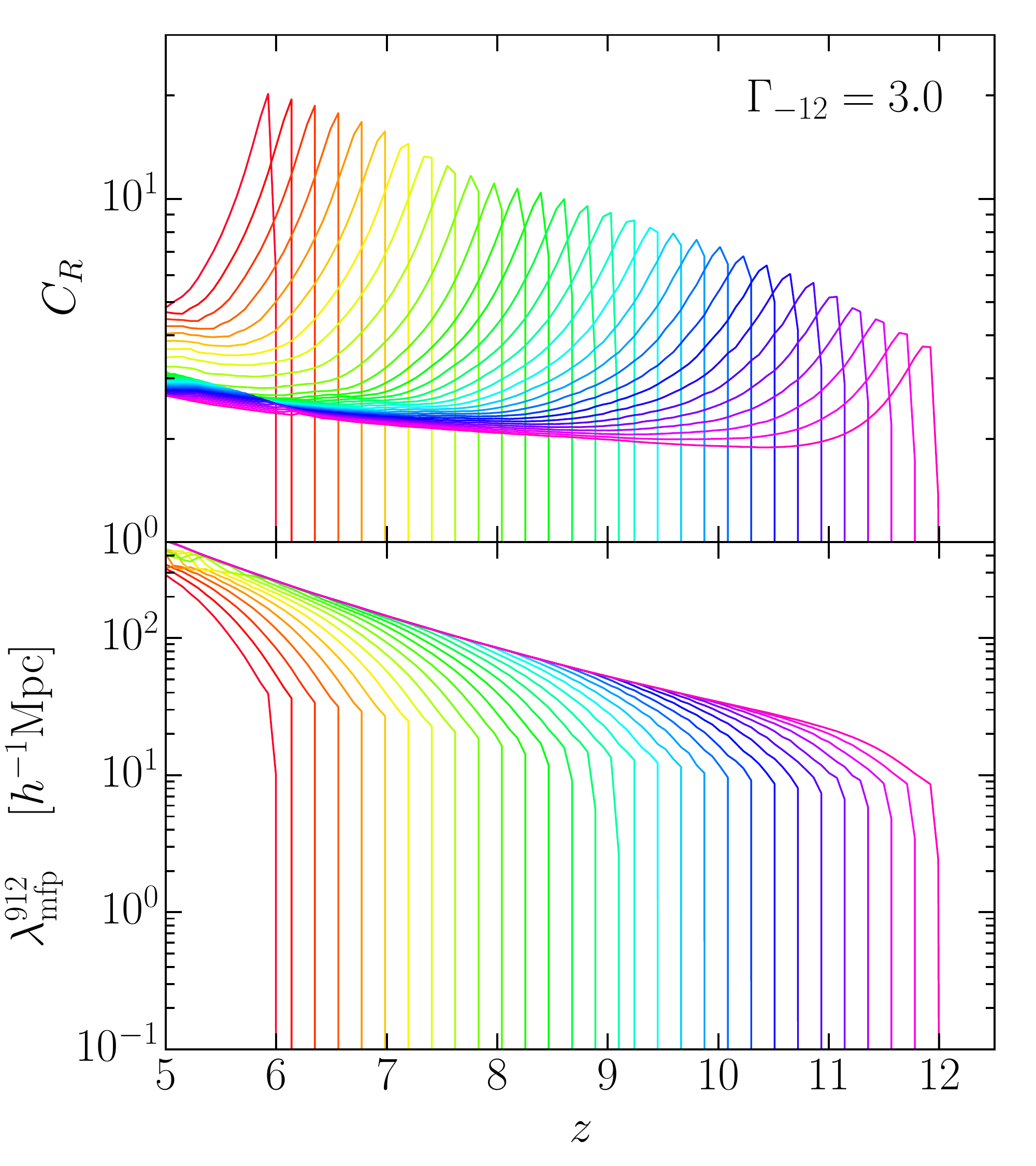} 
\caption{Clumping factors and mean free paths for regions that were reionized at different redshifts.  We obtain these results by interpolating between our $\delta/\sigma = 0$ simulations.  The left and right panels correspond to $\Gamma_{-12}=0.3$ and 3.0, respectively.  In \S \ref{sec:implications} these interpolations are used to construct a simple model for assessing the impact of gas relaxation on the global reionization process.   }
\label{fig:patchy_C_and_MFP}
\end{figure*}

 We now use the ``local" results from our simulation boxes to explore implications for the global reionization process. To motivate our semi-analytic approach, consider how the gas evolves in different locations of the universe: Fig. \ref{fig:patchy_C_and_MFP} shows the clumping factor and mean free path as a function of redshift for 30 values of $\zreion$ and $\delta/\sigma = 0$.  The left and right panels correspond to $\Gamma_{-12}=0.3$ and 3.0, respectively. (To create these plots we have interpolated results between our simulations at $\zreion = 6,8$, and 12.)  At any snapshot in time, patches that were reionized at different times are at different stages of dynamical relaxation.  These panels illustrate the level of spatial variation in $C_R$ and $\MFP$ that result.  For example, at $z=7$, the $C_R$ of recently reionized gas is a factor of $2-3$ higher than of gas reionized at $z>9$. The spatial variations persist well after the last gas parcel has been reionized.    Our aim in this section is to estimate the impact of these relaxation effects on the global reionization process using a simplistic model.

 \begin{figure*}
\includegraphics[width=8.5cm]{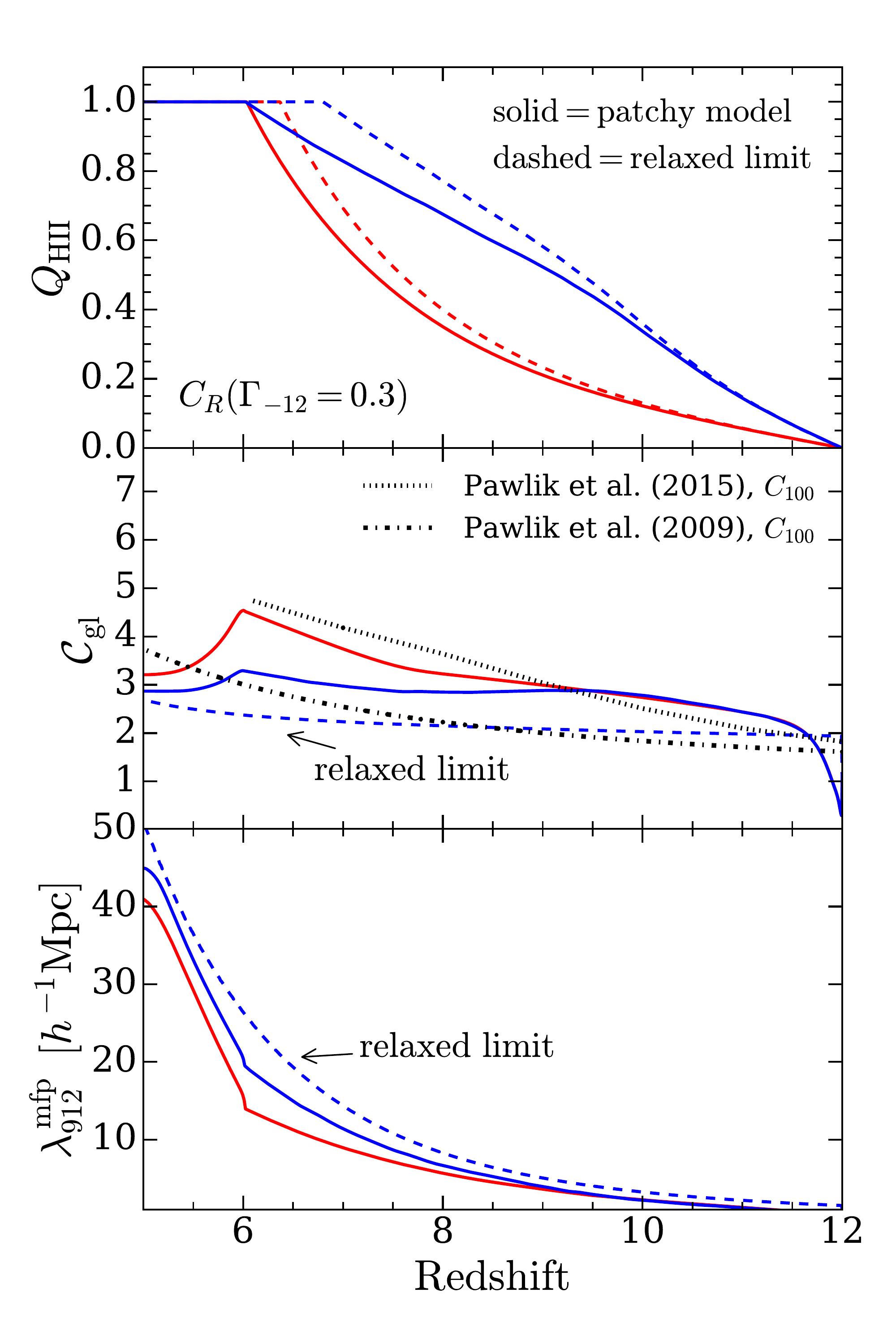} 
\includegraphics[width=8.5cm]{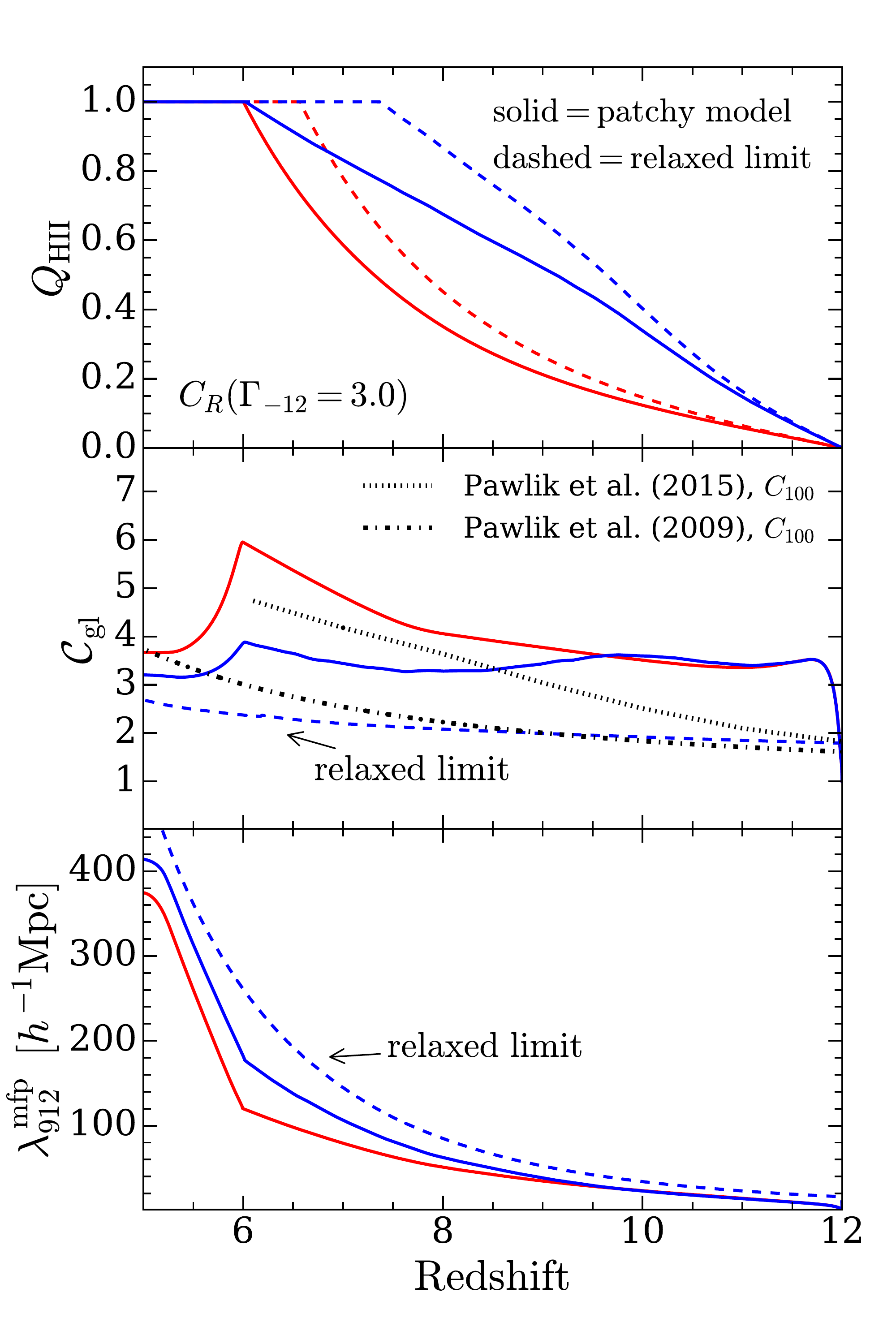} 
\caption{The impact of relaxation on the global reionization process. We use two ionizing emissivities from the literature to explore a rapid (red) and gradual (blue) reionization process.  The top, middle, and bottom panels show the global ionized fraction, clumping factor, and mean free path, respectively.  Our ``patchy" models account for the fact that regions are reionized at different times, and are thus at different stages of relaxation. Ionizing emissivities are re-scaled such that reionization ends at $z=6$ in all the patchy models.  To gauge the effect of relaxation, we compare the models against corresponding relaxed limit models (dashed curves) in which we adopt the relaxed/late-time clumping factor shown as the blue dashed curves in the middle panels (see text for details).  The left and right panels adopt the clumping factors from our simulations with $\Gamma_{-12}=0.3$ and 3.0, respectively.   We find that un-relaxed gaseous structures delay reonization and increase the total number of recombinations by up to a factor of 2.  Rapid reionization leads to a shorter mean free path.      }
\label{fig:G0.3_patchy_model}
\end{figure*}
 
 \subsection{Clumping factor and the ionizing photon budget}
 
 Our starting point is the familiar reionization accounting equation \citep{1987ApJ...321L.107S, 1999ApJ...514..648M}, which has been widely used to model the global ionized fraction ($Q_{\mathrm{HII}}$),
 \begin{equation}
 \frac{d Q_{\mathrm{HII}}}{dt} = \frac{\dot{n}_{\mathrm{ion}}(t)}{\langle n_{\rm H}\rangle} - \frac{\mathcal{R}[Q_{\rm HII}(t)] }{ \langle n_H \rangle},
 \label{eq:reionization_balance}
 \end{equation}
 where $\dot{n}_{\rm ion}$ is the proper ionizing emissivity (the number of ionizing photons per unit time, per unit volume, produced by the sources), and $\langle n_{\rm H} \rangle$ is the mean proper hydrogen number density.  The quantity $\mathcal{R}$ is the volume-weighted mean recombination rate, and the notation $\mathcal{R} = \mathcal{R}[Q_{\rm HII}(t)]$ denotes that the recombination rate is a functional of the global reionization history.  Previous analyses based on equation (\ref{eq:reionization_balance}) have taken $\mathcal{R}$ to be a function of time only, extracted from cosmological simulations.  Here we will take account of its dependence on the reionization history.

 Our model relates the local recombination rates measured in our simulations boxes to the global quantity $\mathcal{R}[Q_{\rm HII}(t)]$. Let us define $R$ to be the local recombination rate, which can be obtained from the clumping factor using $R = C_R~ \alpha_B(T_{\rm ref}) \langle n_e \rangle \langle n_{\rm HII} \rangle$.  In reality $R$ varies from location to location because of spatial variations in the ionizing radiation background ($\Gamma_{-12}$), temperature, density, and local timing of reionization ($\zreion$).  For simplicity, we will focus solely on variations in $\zreion$, fixing $\Gamma_{-12}$ to a constant value and $\delta/\sigma = 0$.    Although neglected here, spatial variations in the ionizing background and their interplay with density fluctuations almost certainly play a significant role in shaping the reionization process. We will gauge the potential impact of background fluctuations by comparing results for different values of $\Gamma_{-12}$.   
 
Under these simplifications, to obtain the global $\mathcal{R}$ we integrate $R(t,\zreion)$ over the probability distribution of $\zreion$, $dP/d \zreion \propto d Q_{\rm HII}/d \zreion$, where the proportionality is set by normalizing to unity.  Plugging this model for $\mathcal{R}$ into equation (\ref{eq:reionization_balance}) yields
\begin{equation}
 \frac{d Q_{\mathrm{HII}}}{dt} = \frac{\dot{n}_{\mathrm{ion}}(t)}{\langle n_{\rm H}\rangle}-\mathcal{C}_{\rm gl}~\alpha_B(T_{\rm ref}) Q_{\rm HII} \langle n_H \rangle (1+\chi),
\label{eq:reionization_balance_2}
\end{equation}   
where we have defined the {\it global} clumping factor,

\begin{equation}
 \mathcal{C}_{\rm gl}[Q_{\rm HII}(t)] =   \int^\infty_{z(t)} dz' \frac{d P}{dz'} C_R(t,z'),
\label{eq:global_clumping}
\end{equation}
as a volume-weighted averge of $C_R$, and we have used that $\langle n_e \rangle \approx \langle n_H \rangle (1+\chi)$, with $\chi \equiv n_{\rm He}/n_{\rm H} = 0.083$ for singly ionized helium, and $Q_{\rm HII} \equiv n_{\rm HII}/n_{\rm H}$.  Note that the last term in equation (\ref{eq:reionization_balance_2}) depends on the reionization history through the integral in $\mathcal{C}_{\rm gl}$.  This makes it slightly more complicated than the standard case but the equation can nonetheless be solved easily using elementary numerical techniques such as Euler's method. 

We adopt the ionizing emissivities from \citet{2015ApJ...802L..19R} and \citet{2019ApJ...879...36F} (specfically their ``minimal AGN" model) to explore illustrative examples of ``rapid" and ``gradual" reionization models, respectively. We fix the start of reionization at $z=12$, since we did not run simulations with $\zreion > 12$, and we rescale the emissivities such that our fiducial models (denoted ``patchy model" below) end at $z\approx 6$.  We evaluate the integral (\ref{eq:global_clumping}) by interpolating $C_R(t,\zreion)$ in two-dimensions using our three runs with $\zreion = 6, 8$, and 12.  The interpolations are illustrated in Fig. \ref{fig:patchy_C_and_MFP}.
 
To gauge the impact of relaxation, we compare our patchy model against a ``relaxed limit" in which all of the gas has relaxed down to the limiting evolution.  For the relaxed models we use $C_R$ from the $\zreion=12$ simulations for $z < 9$.  Above this redshift, we extrapolate backward in time using a power-law fit of the regime $8<z<9$.  Fitting to this redshift range is motivated by visual inspection, but our analysis is insensitive to the details of the extrapolation because the relaxed limit of $C_R$ is very weakly dependent on redshift (see e.g. Fig. \ref{fig:G0.3_patchy_model}).
 
 The solid curves in the left (right) panel of Fig. \ref{fig:G0.3_patchy_model} show the results of our patchy model with $\Gamma_{-12} = 0.3$ (3.0).  We adopt the former as our fiducial value because both $\Gamma_{-12}$ and the derived global $\MFP$ (which we will discuss below) at $z<6$ are more consistent with observational measurements \citep[e.g.][]{2018MNRAS.473..560D,2014MNRAS.445.1745W}.  The top and middle panels show the global ionized fraction and clumping factor, respectively.  The red and blue curves correspond to the rapid and gradual scenarios. The relaxed limit clumping factors are shown as the blue/dashed curves in the middle panels of Fig. \ref{fig:G0.3_patchy_model}, and the corresponding reionization histories for the rapid and gradual scenarios are shown as red and blue dashed curves in the top panels.\footnote{The relaxed models (dashed) adopt the same ionizing emissivities as the correpsonding patchy models (solid curves).} The un-relaxed gas, in which recombination rates are higher, causes a significant delay in the global reionization history. This can be seen in the top panel by comparing the dashed curves to their respective solid curves. 
 
 The evolution of the global clumping factors is shown in the middle panels of Fig. \ref{fig:G0.3_patchy_model}.   Near the beginning of reionization, essentially all ionized gas is un-relaxed. As reionization progresses, the evolution of $\mathcal{C}_{\rm gl}$ is set by a competition between the un-relaxed, freshly ionized gas, and the relaxed gas that was reionized long ago.   At the end of reionization $\mathcal{C}_{\rm gl}$ begins to fall because the production of un-relaxed, clumpy gas halts. Notably, we find that $\mathcal{C}_{\rm gl}$ depends somewhat on the reionization history.  Intuitively, if more volume is reionized later (such as in our rapid model), the clumping factor rises to a larger value.   
 
 For comparison, we also show a couple of clumping factors from the literature.  The dotted curve in the middle panel of Fig. \ref{fig:G0.3_patchy_model} is from the L25N512 simulation of \citet{2015MNRAS.451.1586P}, which was used in the recent study of \citet{2019ApJ...879...36F}.  The dot-dashed curve shows the form $C_{100} = 1 + 43z^{-1.71}$ from \citet{2009MNRAS.394.1812P}, which was adopted in the widely used model of \citet{2012ApJ...746..125H} for the ionizing radiation background.  They define $C_{100}\equiv \langle \rho_b^2 \rangle_{100}/\langle \rho_b \rangle$, where $\rho_b$ is the baryonic matter density, and $\langle \ldots \rangle_{100}$ denotes an average over gas with overdensities $\leq 100$. In spite of this and other major differences in the numerical methods and approximations (e.g. our simplistic model adopts constant values of $\Gamma_{-12}$), our results are reasonably similar to the most recently published clumping factors. The main qualitative difference is that $\mathcal{C}_{\rm gl}$ drops off after the last gas parcels are reionized, in a manner that depends on the reionization history.

 Un-relaxed gas contributes significantly to the total ionizing photon budget. A comparison between the number of ionizing photons per hydrogen atom required to complete  reionization, $N_{\rm ion} = \int dt~\dot{n}_{\rm ion} / \langle n_{\rm H} \rangle$, supports this conclusion.   Using our $\Gamma_{-12}=0.3$ model, the relaxed limits of the rapid and gradual scenarios require 1.47 and 1.72 photons per hydrogen atom, respectively, to complete reionization.  Comparing these numbers to the higher cost of 1.80 and 2.12 photons per hydrogen atom in the corresponding patchy models indicates a $50-70$ \% boost to total number of recombinations ($N_{\rm ion} -1 $) due to un-relaxed gas.\footnote{For $\Gamma_{-12}=3.0$ the recombination rate is boosted by 70 and 240 \% in the rapid and gradual scenarios, respectively. However, we note that $\Gamma_{-12}=3.0$ is unrealistically large for the mean photoionization rate during reionization.}

 Lastly, the bottom panels of Fig. \ref{fig:G0.3_patchy_model} consider the evolution of the comoving mean free path averaged over ionized regions. Similar to eq. (\ref{eq:global_clumping}), we define the global mean free path in ionized gas by an average over local absorption coefficients, $\mathcal{\lambda}^{-1}_{\rm gl}(z) \equiv \int_{z}^{\infty}~dz~(\MFP)^{-1} dP/dz'$. The relaxed limits (dashed) were extrapolated to high redshift in a manner similar to the clumping factor.  A more rapid reionization process leads to a modestly shorter post-reionization mean free path owing to the effects of relaxation.  This is evident in the comparison between our models at $z<6$. At $z=5.8~(5.2)$, the mean free path in our rapid model is $20~(10)\%$ shorter than in the gradual one (using the $\Gamma_{-12}=0.3$ case).

 \subsection{Spatial fluctuations in the mean free path}

 Figure \ref{fig:patchy_C_and_MFP} shows that the relaxation process naturally leads to spatial variations in the mean free path.  During reionization, the mean free path in recently reionized gas can be a factor of a few lower compared to relaxed regions.  Even after reionization is over, spatial variations persist for the relaxation timescale of a few hundred million years. Consider, for example, the bottom-left panel of Fig. \ref{fig:patchy_C_and_MFP}.  If the last neutral gas parcels are reionized at $z=6$, relaxation alone leads to a factor of 2-3 variation in the mean free path at $z=5.5$.  
 
 This may have observable consequences for the $z>5$ Ly$\alpha$ forest. It has been proposed that the $160$ Mpc-long Ly$\alpha$ trough towards quasar ULAS J0148+0600 \citep{2015MNRAS.447.3402B} may owe to neutral islands intersecting the sight line \citep{2020MNRAS.491.1736K, 2019arXiv191003570N}.  If true, the ionized regions bordering the neutral islands were likely in the process of relaxing at the epoch of observation. In this case, the un-relaxed gas could contribute to the suppression of $\Gamma_{-12}$ in the viscinity of the neutral islands, and the enhanced clumpiness may produce signatures in the small-scale structure of the forest.  These possible effects are absent in the post-processed RT and semi-numerical simulations used thus far to explore late-ending reionization scenarios \citep[e.g.][]{2019MNRAS.485L..24K, 2020MNRAS.491.1736K, 2019arXiv191003570N}.   It is furthermore unclear to what degree relaxation is captured in large-volume, fully coupled RT simulations because of the extreme dynamic range required to resolve it. Current models of reionization may be missing important effects from hydrodynamic relaxation.

 \section{Conclusion}
 \label{sec:conclusion}
 
 There have been relatively few studies of the clumpiness of the recently-ionized IGM relative to the properties of the sources that drove reionization.  This is despite the modern understanding that the clumpiness -- which sets the mean free path in ionized regions -- shapes the structure of reionization as much as the sources \citep{2005MNRAS.363.1031F, 2007MNRAS.377.1043M, 2012ApJ...747..126A}.  Additionally, constraints on the global rate of ionizing emissions are limited by our understanding of the mean free path \citep{2007MNRAS.382..325B,2013MNRAS.436.1023B,2018MNRAS.473..560D}.  Previous studies of clumpiness have primarily focused on the relaxed limit well after ionization \citep[e.g.][]{2000ApJ...530....1M, 2009MNRAS.394.1812P, 2012ApJ...747..100S, 2012arXiv1209.2489F}, and those that followed the hydrodynamic response did so with prescriptions for the radiation field or did not demonstrate convergence in resolution nor boxsize \citep{2009MNRAS.394.1812P, 2015ApJ...810..154K,2018MNRAS.478.5123R, 2016ApJ...831...86P}.

We presented a systematic exploration of the clumpiness of the recently ionized IGM in the concordance $\Lambda$CDM cosmology.  This exploration used extremely high-resolution ray tracing radiative transfer that captured the $\sim 10^4M_\odot$ Jeans scale for an unheated IGM, combined with DC-mode simulations to explore the effect of long wavelength modes. We presented a new Gauss-Hermite quadrature approach for averaging over DC-mode simulations, which should be significantly more accurate than the simpler summation methods of prior studies.  Our simulations reveal a complex ionization and heating process in which the smallest structures initially self-shield but are ultimately photoevaporated, driving a hydrodynamic response into the relaxing, ionized IGM (c.f. Fig.~\ref{fig:intro}). The sound waves are initialized at the time of ionization, leading to a coherent acoustic scale in the gas density in regions reionized at the same time.

We found that the relaxation of reionized gas lasts hundreds of millions of years.  The clumping factor (which is proportional to the recombination rate) of the ionized gas can reach as high as $10-20$ for regions reionized near $z\sim 6$, close to the likely end of reionization, with the peak value depending on the incident ionizing intensity and the degree to which the gas had been pre-heating by X-ray sources.  The peak clumping occurs $10~$Myr after ionization, set by a balance between how deep the I-fronts penetrate into dense regions and the time-scale of their relaxation.  The clumping factor decays to roughly half its peak value $100~$Myr later as the gas responds to the photoheating, reaching its final relaxed value of $\approx 3$ after $300~$Myr.  Because $300~$Myr is a substantial fraction of reionization's duration, differences in the local timing of ionization introduce spatial variations in the clumpiness of the IGM.  

To understand the impact of un-relaxed gas on the global reionization history, we implemented our time-dependent clumping factors into a simple semi-analytic model that accounts for the inhomogeneity of $\zreion$.  We found that un-relaxed gas boosts the mean clumping factor during reionization by a factor of 1.5-1.7, leading to a similar enhacement in the total number of recombinations. This boost can delay reionization by $\Delta z = 0.4-0.8$, with a longer delay for more extended reionization histories.  Our prescription for treating the inhomogeneous gas clumping can be similarly applied in future studies assessing the ability of the observed galaxy population to reionize the Universe.

The other side of the coin to gas clumpiness is the mean free path in ionized regions.  The mean free path evolves in unison with the clumping factor, reaching its relaxed evolution of $\MFP \propto (1+z)^{-4.4}$ (at fixed $\Gamma_{-12}$) within a few hundred million years.  Notably, this is also the evolution of the (comoving) mean free path measured observationally at $z \lesssim 5.2$ by \citet{2014MNRAS.445.1745W}.  The long timescale of relaxation has implications for the mean free path and its spatial fluctuations.  A more gradual reionization process, in which more volume is reionized at earlier times, leads to a longer post-reionization mean free path.  The mean free path is also more inhomogeneous because the local values in regions that were reionized more recently can be substantially shorter than in those that were reionized longer ago.        

Our results for the clumping factor and mean free path are well-suited to be applied in sub-grid models for both semi-numeric and radiative transfer reionization simulation codes.  In light of the emerging picture in which the $z>6$ forest may be probing deep into reionization, another topic of interest is how relaxation manifests in the forest.   There is evidence that some high-redshift quasars were active for $\lesssim 10$ Myr at the epoch of observation, and were shining into a largely neutral medium \citep[e.g.][]{2017ApJ...840...24E, 2019ApJ...884L..19D}. Our results imply that these proximity zones may contain un-relaxed gas.  A preliminary investigation by us suggests that the fine-grained clumpiness of this un-relaxed gas is largely masked by the  considerable thermal broadening of gas recently ionized by a hard and intense quasar spectrum.  Outside of proximity zones, forest transmission peaks arising from hot, recently ionized regions may also be a signpost for un-relaxed gas.  However, un-relaxed regions also correspond to shorter mean free paths, and thus more absorption from attenuation of the local ionizing background.  The potential signatures of recently ionized gas in the Ly$\alpha$ forest merits further study.

\acknowledgements
The authors thank the anonymous referee and Fred Davies for helpful comments on this manuscript.  This research was supported by HST award HST-AR15013.005-A. M.M. acknowledges support from NSF grants AST~1312724 and AST~1614439 and NASA grant ATP~
NNX17AH68G. A.M acknowledges support from the European Research Countil (ERC) through a Starting Grant award AIDA (grant no. 638809). All computations were performed with NSF
XSEDE allocation TG-AST120066.

\bibliographystyle{apj}
\bibliography{master}

\begin{thebibliography}{}
\expandafter\ifx\csname natexlab\endcsname\relax\def\natexlab#1{#1}\fi

\bibitem[{{Altay} {et~al.}(2011){Altay}, {Theuns}, {Schaye}, {Crighton}, \&
  {Dalla Vecchia}}]{2011ApJ...737L..37A}
{Altay}, G., {Theuns}, T., {Schaye}, J., {Crighton}, N. H.~M., \& {Dalla
  Vecchia}, C. 2011, \apjl, 737, L37

\bibitem[{{Alvarez} \& {Abel}(2012)}]{2012ApJ...747..126A}
{Alvarez}, M.~A., \& {Abel}, T. 2012, \apj, 747, 126

\bibitem[{{Ba{\~n}ados} {et~al.}(2018){Ba{\~n}ados}, {Venemans},
  {Mazzucchelli}, {Farina}, {Walter}, {Wang}, {Decarli}, {Stern}, {Fan},
  {Davies}, {Hennawi}, {Simcoe}, {Turner}, {Rix}, {Yang}, {Kelson}, {Rudie}, \&
  {Winters}}]{2018Natur.553..473B}
{Ba{\~n}ados}, E., {Venemans}, B.~P., {Mazzucchelli}, C., {et~al.} 2018, \nat,
  553, 473

\bibitem[{{Becker} \& {Bolton}(2013)}]{2013MNRAS.436.1023B}
{Becker}, G.~D., \& {Bolton}, J.~S. 2013, \mnras, 436, 1023

\bibitem[{{Becker} {et~al.}(2015){Becker}, {Bolton}, {Madau}, {Pettini},
  {Ryan-Weber}, \& {Venemans}}]{2015MNRAS.447.3402B}
{Becker}, G.~D., {Bolton}, J.~S., {Madau}, P., {et~al.} 2015, \mnras, 447, 3402

\bibitem[{{Bolton} \& {Haehnelt}(2007)}]{2007MNRAS.382..325B}
{Bolton}, J.~S., \& {Haehnelt}, M.~G. 2007, \mnras, 382, 325

\bibitem[{{Bouwens} {et~al.}(2015){Bouwens}, {Illingworth}, {Oesch}, {Trenti},
  {Labb{\'e}}, {Bradley}, {Carollo}, {van Dokkum}, {Gonzalez}, {Holwerda},
  {Franx}, {Spitler}, {Smit}, \& {Magee}}]{2015ApJ...803...34B}
{Bouwens}, R.~J., {Illingworth}, G.~D., {Oesch}, P.~A., {et~al.} 2015, \apj,
  803, 34

\bibitem[{{Chardin} {et~al.}(2018){Chardin}, {Kulkarni}, \&
  {Haehnelt}}]{2018MNRAS.478.1065C}
{Chardin}, J., {Kulkarni}, G., \& {Haehnelt}, M.~G. 2018, \mnras, 478, 1065

\bibitem[{{Choudhury} {et~al.}(2009){Choudhury}, {Haehnelt}, \&
  {Regan}}]{2009MNRAS.394..960C}
{Choudhury}, T.~R., {Haehnelt}, M.~G., \& {Regan}, J. 2009, \mnras, 394, 960

\bibitem[{{Crociani} {et~al.}(2011){Crociani}, {Mesinger}, {Moscardini}, \&
  {Furlanetto}}]{2011MNRAS.411..289C}
{Crociani}, D., {Mesinger}, A., {Moscardini}, L., \& {Furlanetto}, S. 2011,
  \mnras, 411, 289

\bibitem[{{D'Aloisio} {et~al.}(2018){D'Aloisio}, {McQuinn}, {Davies}, \&
  {Furlanetto}}]{2018MNRAS.473..560D}
{D'Aloisio}, A., {McQuinn}, M., {Davies}, F.~B., \& {Furlanetto}, S.~R. 2018,
  \mnras, 473, 560

\bibitem[{{D'Aloisio} {et~al.}(2019){D'Aloisio}, {McQuinn}, {Maupin}, {Davies},
  {Trac}, {Fuller}, \& {Upton Sanderbeck}}]{2019ApJ...874..154D}
{D'Aloisio}, A., {McQuinn}, M., {Maupin}, O., {et~al.} 2019, \apj, 874, 154

\bibitem[{{Davies} \& {Furlanetto}(2016)}]{2015arXiv150907131D}
{Davies}, F.~B., \& {Furlanetto}, S.~R. 2016, \mnras, 460, 1328

\bibitem[{{Davies} {et~al.}(2019){Davies}, {Hennawi}, \&
  {Eilers}}]{2019ApJ...884L..19D}
{Davies}, F.~B., {Hennawi}, J.~F., \& {Eilers}, A.-C. 2019, \apjl, 884, L19

\bibitem[{{Davies} {et~al.}(2018){Davies}, {Hennawi}, {Ba{\~n}ados},
  {Luki{\'c}}, {Decarli}, {Fan}, {Farina}, {Mazzucchelli}, {Rix}, {Venemans},
  {Walter}, {Wang}, \& {Yang}}]{2018ApJ...864..142D}
{Davies}, F.~B., {Hennawi}, J.~F., {Ba{\~n}ados}, E., {et~al.} 2018, \apj, 864,
  142

\bibitem[{{Deparis} {et~al.}(2019){Deparis}, {Aubert}, {Ocvirk}, {Chardin}, \&
  {Lewis}}]{2019A&A...622A.142D}
{Deparis}, N., {Aubert}, D., {Ocvirk}, P., {Chardin}, J., \& {Lewis}, J. 2019,
  \aap, 622, A142

\bibitem[{{Doussot} {et~al.}(2019){Doussot}, {Trac}, \&
  {Cen}}]{2019ApJ...870...18D}
{Doussot}, A., {Trac}, H., \& {Cen}, R. 2019, \apj, 870, 18

\bibitem[{{Eilers} {et~al.}(2017){Eilers}, {Davies}, {Hennawi}, {Prochaska},
  {Luki{\'c}}, \& {Mazzucchelli}}]{2017ApJ...840...24E}
{Eilers}, A.-C., {Davies}, F.~B., {Hennawi}, J.~F., {et~al.} 2017, \apj, 840,
  24

\bibitem[{{Emberson} {et~al.}(2013){Emberson}, {Thomas}, \&
  {Alvarez}}]{2013ApJ...763..146E}
{Emberson}, J.~D., {Thomas}, R.~M., \& {Alvarez}, M.~A. 2013, \apj, 763, 146

\bibitem[{{Fan} {et~al.}(2006){Fan}, {Strauss}, {Becker}, {White}, {Gunn},
  {Knapp}, {Richards}, {Schneider}, {Brinkmann}, \& {Fukugita}}]{fan06}
{Fan}, X., {Strauss}, M.~A., {Becker}, R.~H., {et~al.} 2006, \aj, 132, 117

\bibitem[{{Fialkov} {et~al.}(2014){Fialkov}, {Barkana}, \&
  {Visbal}}]{2014Natur.506..197F}
{Fialkov}, A., {Barkana}, R., \& {Visbal}, E. 2014, \nat, 506, 197

\bibitem[{{Finkelstein} {et~al.}(2019){Finkelstein}, {D'Aloisio},
  {Paardekooper}, {Ryan}, {Behroozi}, {Finlator}, {Livermore}, {Upton
  Sanderbeck}, {Dalla Vecchia}, \& {Khochfar}}]{2019ApJ...879...36F}
{Finkelstein}, S.~L., {D'Aloisio}, A., {Paardekooper}, J.-P., {et~al.} 2019,
  \apj, 879, 36

\bibitem[{{Finlator} {et~al.}(2012){Finlator}, {Oh}, {{\"O}zel}, \&
  {Dav{\'e}}}]{2012arXiv1209.2489F}
{Finlator}, K., {Oh}, S.~P., {{\"O}zel}, F., \& {Dav{\'e}}, R. 2012, ArXiv
  e-prints, arXiv:1209.2489

\bibitem[{{Furlanetto}(2006)}]{2006MNRAS.371..867F}
{Furlanetto}, S.~R. 2006, \mnras, 371, 867

\bibitem[{{Furlanetto} \& {Oh}(2005)}]{2005MNRAS.363.1031F}
{Furlanetto}, S.~R., \& {Oh}, S.~P. 2005, \mnras, 363, 1031

\bibitem[{{Gnedin}(2014)}]{2014ApJ...793...29G}
{Gnedin}, N.~Y. 2014, \apj, 793, 29

\bibitem[{{Gnedin} {et~al.}(2011){Gnedin}, {Kravtsov}, \&
  {Rudd}}]{2011ApJS..194...46G}
{Gnedin}, N.~Y., {Kravtsov}, A.~V., \& {Rudd}, D.~H. 2011, \apjs, 194, 46

\bibitem[{{Greig} {et~al.}(2017){Greig}, {Mesinger}, {Haiman}, \&
  {Simcoe}}]{2017MNRAS.466.4239G}
{Greig}, B., {Mesinger}, A., {Haiman}, Z., \& {Simcoe}, R.~A. 2017, \mnras,
  466, 4239

\bibitem[{{Haardt} \& {Madau}(2012)}]{2012ApJ...746..125H}
{Haardt}, F., \& {Madau}, P. 2012, \apj, 746, 125

\bibitem[{{Iliev} {et~al.}(2014){Iliev}, {Mellema}, {Ahn}, {Shapiro}, {Mao}, \&
  {Pen}}]{2014MNRAS.439..725I}
{Iliev}, I.~T., {Mellema}, G., {Ahn}, K., {et~al.} 2014, \mnras, 439, 725

\bibitem[{{Iliev} {et~al.}(2005){Iliev}, {Shapiro}, \&
  {Raga}}]{2005MNRAS.361..405I}
{Iliev}, I.~T., {Shapiro}, P.~R., \& {Raga}, A.~C. 2005, \mnras, 361, 405

\bibitem[{{Ir{\v{s}}i{\v{c}}} {et~al.}(2017){Ir{\v{s}}i{\v{c}}}, {Viel},
  {Haehnelt}, {Bolton}, {Cristiani}, {Becker}, {D'Odorico}, {Cupani}, {Kim},
  {Berg}, {L{\'o}pez}, {Ellison}, {Christensen}, {Denney}, \&
  {Worseck}}]{2017PhRvD..96b3522I}
{Ir{\v{s}}i{\v{c}}}, V., {Viel}, M., {Haehnelt}, M.~G., {et~al.} 2017, \prd,
  96, 023522

\bibitem[{{Kashikawa} {et~al.}(2006){Kashikawa}, {Shimasaku}, {Malkan}, {Doi},
  {Matsuda}, {Ouchi}, {Taniguchi}, {Ly}, {Nagao}, {Iye}, {Motohara},
  {Murayama}, {Murozono}, {Nariai}, {Ohta}, {Okamura}, {Sasaki}, {Shioya}, \&
  {Umemura}}]{2006ApJ...648....7K}
{Kashikawa}, N., {Shimasaku}, K., {Malkan}, M.~A., {et~al.} 2006, \apj, 648, 7

\bibitem[{{Kaurov} \& {Gnedin}(2015)}]{2015ApJ...810..154K}
{Kaurov}, A.~A., \& {Gnedin}, N.~Y. 2015, \apj, 810, 154

\bibitem[{{Keating} {et~al.}(2020){Keating}, {Weinberger}, {Kulkarni},
  {Haehnelt}, {Chardin}, \& {Aubert}}]{2020MNRAS.491.1736K}
{Keating}, L.~C., {Weinberger}, L.~H., {Kulkarni}, G., {et~al.} 2020, \mnras,
  491, 1736

\bibitem[{{Kulkarni} {et~al.}(2019{\natexlab{a}}){Kulkarni}, {Keating},
  {Haehnelt}, {Bosman}, {Puchwein}, {Chardin}, \&
  {Aubert}}]{2019MNRAS.485L..24K}
{Kulkarni}, G., {Keating}, L.~C., {Haehnelt}, M.~G., {et~al.}
  2019{\natexlab{a}}, \mnras, 485, L24

\bibitem[{{Kulkarni} {et~al.}(2019{\natexlab{b}}){Kulkarni}, {Worseck}, \&
  {Hennawi}}]{2019MNRAS.488.1035K}
{Kulkarni}, G., {Worseck}, G., \& {Hennawi}, J.~F. 2019{\natexlab{b}}, \mnras,
  488, 1035

\bibitem[{Lewis {et~al.}(2000)Lewis, Challinor, \& Lasenby}]{Lewis:1999bs}
Lewis, A., Challinor, A., \& Lasenby, A. 2000, Astrophys. J., 538, 473

\bibitem[{{Madau} {et~al.}(1999){Madau}, {Haardt}, \&
  {Rees}}]{1999ApJ...514..648M}
{Madau}, P., {Haardt}, F., \& {Rees}, M.~J. 1999, \apj, 514, 648

\bibitem[{{McQuinn}(2016)}]{2016ARA&A..54..313M}
{McQuinn}, M. 2016, \araa, 54, 313

\bibitem[{{McQuinn} {et~al.}(2007){McQuinn}, {Lidz}, {Zahn}, {Dutta},
  {Hernquist}, \& {Zaldarriaga}}]{2007MNRAS.377.1043M}
{McQuinn}, M., {Lidz}, A., {Zahn}, O., {et~al.} 2007, \mnras, 377, 1043

\bibitem[{{McQuinn} {et~al.}(2011){McQuinn}, {Oh}, \&
  {Faucher-Gigu{\`e}re}}]{2011ApJ...743...82M}
{McQuinn}, M., {Oh}, S.~P., \& {Faucher-Gigu{\`e}re}, C.-A. 2011, \apj, 743, 82

\bibitem[{{McQuinn} \& {Upton Sanderbeck}(2016)}]{2016MNRAS.456...47M}
{McQuinn}, M., \& {Upton Sanderbeck}, P.~R. 2016, \mnras, 456, 47

\bibitem[{{Mesinger} {et~al.}(2015){Mesinger}, {Aykutalp}, {Vanzella},
  {Pentericci}, {Ferrara}, \& {Dijkstra}}]{2015MNRAS.446..566M}
{Mesinger}, A., {Aykutalp}, A., {Vanzella}, E., {et~al.} 2015, \mnras, 446, 566

\bibitem[{{Mesinger} \& {Furlanetto}(2009)}]{2009MNRAS.400.1461M}
{Mesinger}, A., \& {Furlanetto}, S. 2009, \mnras, 400, 1461

\bibitem[{{Miralda-Escud{\'e}} {et~al.}(2000){Miralda-Escud{\'e}}, {Haehnelt},
  \& {Rees}}]{2000ApJ...530....1M}
{Miralda-Escud{\'e}}, J., {Haehnelt}, M., \& {Rees}, M.~J. 2000, \apj, 530, 1

\bibitem[{{Mortlock} {et~al.}(2011){Mortlock}, {Warren}, {Venemans}, {Patel},
  {Hewett}, {McMahon}, {Simpson}, {Theuns}, {Gonz{\'a}les-Solares}, {Adamson},
  {Dye}, {Hambly}, {Hirst}, {Irwin}, {Kuiper}, {Lawrence}, \&
  {R{\"o}ttgering}}]{mortlock11}
{Mortlock}, D.~J., {Warren}, S.~J., {Venemans}, B.~P., {et~al.} 2011, \nat,
  474, 616

\bibitem[{{Nasir} \& {D'Aloisio}(2019)}]{2019arXiv191003570N}
{Nasir}, F., \& {D'Aloisio}, A. 2019, arXiv e-prints, arXiv:1910.03570

\bibitem[{{Ocvirk} {et~al.}(2016){Ocvirk}, {Gillet}, {Shapiro}, {Aubert},
  {Iliev}, {Teyssier}, {Yepes}, {Choi}, {Sullivan}, {Knebe}, {Gottl{\"o}ber},
  {D'Aloisio}, {Park}, {Hoffman}, \& {Stranex}}]{2016MNRAS.463.1462O}
{Ocvirk}, P., {Gillet}, N., {Shapiro}, P.~R., {et~al.} 2016, \mnras, 463, 1462

\bibitem[{{Ono} {et~al.}(2012){Ono}, {Ouchi}, {Mobasher}, {Dickinson},
  {Penner}, {Shimasaku}, {Weiner}, {Kartaltepe}, {Nakajima}, {Nayyeri},
  {Stern}, {Kashikawa}, \& {Spinrad}}]{ono12}
{Ono}, Y., {Ouchi}, M., {Mobasher}, B., {et~al.} 2012, \apj, 744, 83

\bibitem[{{Park} {et~al.}(2016){Park}, {Shapiro}, {Choi}, {Yoshida}, {Hirano},
  \& {Ahn}}]{2016ApJ...831...86P}
{Park}, H., {Shapiro}, P.~R., {Choi}, J.-h., {et~al.} 2016, \apj, 831, 86

\bibitem[{{Pawlik} {et~al.}(2015){Pawlik}, {Schaye}, \& {Dalla
  Vecchia}}]{2015MNRAS.451.1586P}
{Pawlik}, A.~H., {Schaye}, J., \& {Dalla Vecchia}, C. 2015, \mnras, 451, 1586

\bibitem[{{Pawlik} {et~al.}(2009){Pawlik}, {Schaye}, \& {van
  Scherpenzeel}}]{2009MNRAS.394.1812P}
{Pawlik}, A.~H., {Schaye}, J., \& {van Scherpenzeel}, E. 2009, \mnras, 394,
  1812

\bibitem[{{Pentericci} {et~al.}(2014){Pentericci}, {Vanzella}, {Fontana},
  {Castellano}, {Treu}, {Mesinger}, {Dijkstra}, {Grazian}, {Brada{\v c}},
  {Conselice}, {Cristiani}, {Dunlop}, {Galametz}, {Giavalisco}, {Giallongo},
  {Koekemoer}, {McLure}, {Maiolino}, {Paris}, \&
  {Santini}}]{2014ApJ...793..113P}
{Pentericci}, L., {Vanzella}, E., {Fontana}, A., {et~al.} 2014, \apj, 793, 113

\bibitem[{{Planck Collaboration} {et~al.}(2018){Planck Collaboration},
  {Aghanim}, {Akrami}, {Ashdown}, {Aumont}, {Baccigalupi}, {Ballardini},
  {Banday}, {Barreiro}, {Bartolo}, {Basak}, {Battye}, {Benabed}, {Bernard},
  {Bersanelli}, {Bielewicz}, {Bock}, {Bond}, {Borrill}, {Bouchet}, {Boulanger},
  {Bucher}, {Burigana}, {Butler}, {Calabrese}, {Cardoso}, {Carron},
  {Challinor}, {Chiang}, {Chluba}, {Colombo}, {Combet}, {Contreras}, {Crill},
  {Cuttaia}, {de Bernardis}, {de Zotti}, {Delabrouille}, {Delouis}, {Di
  Valentino}, {Diego}, {Dor{\'e}}, {Douspis}, {Ducout}, {Dupac}, {Dusini},
  {Efstathiou}, {Elsner}, {En{\ss}lin}, {Eriksen}, {Fantaye}, {Farhang},
  {Fergusson}, {Fernandez-Cobos}, {Finelli}, {Forastieri}, {Frailis},
  {Fraisse}, {Franceschi}, {Frolov}, {Galeotta}, {Galli}, {Ganga},
  {G{\'e}nova-Santos}, {Gerbino}, {Ghosh}, {Gonz{\'a}lez-Nuevo}, {G{\'o}rski},
  {Gratton}, {Gruppuso}, {Gudmundsson}, {Hamann}, {Handley}, {Hansen},
  {Herranz}, {Hildebrandt}, {Hivon}, {Huang}, {Jaffe}, {Jones}, {Karakci},
  {Keih{\"a}nen}, {Keskitalo}, {Kiiveri}, {Kim}, {Kisner}, {Knox},
  {Krachmalnicoff}, {Kunz}, {Kurki-Suonio}, {Lagache}, {Lamarre}, {Lasenby},
  {Lattanzi}, {Lawrence}, {Le Jeune}, {Lemos}, {Lesgourgues}, {Levrier},
  {Lewis}, {Liguori}, {Lilje}, {Lilley}, {Lindholm}, {L{\'o}pez-Caniego},
  {Lubin}, {Ma}, {Mac{\'\i}as-P{\'e}rez}, {Maggio}, {Maino}, {Mandolesi},
  {Mangilli}, {Marcos-Caballero}, {Maris}, {Martin}, {Martinelli},
  {Mart{\'\i}nez-Gonz{\'a}lez}, {Matarrese}, {Mauri}, {McEwen}, {Meinhold},
  {Melchiorri}, {Mennella}, {Migliaccio}, {Millea}, {Mitra},
  {Miville-Desch{\^e}nes}, {Molinari}, {Montier}, {Morgante}, {Moss}, {Natoli},
  {N{\o}rgaard-Nielsen}, {Pagano}, {Paoletti}, {Partridge}, {Patanchon},
  {Peiris}, {Perrotta}, {Pettorino}, {Piacentini}, {Polastri}, {Polenta},
  {Puget}, {Rachen}, {Reinecke}, {Remazeilles}, {Renzi}, {Rocha}, {Rosset},
  {Roudier}, {Rubi{\~n}o-Mart{\'\i}n}, {Ruiz-Granados}, {Salvati}, {Sandri},
  {Savelainen}, {Scott}, {Shellard}, {Sirignano}, {Sirri}, {Spencer},
  {Sunyaev}, {Suur-Uski}, {Tauber}, {Tavagnacco}, {Tenti}, {Toffolatti},
  {Tomasi}, {Trombetti}, {Valenziano}, {Valiviita}, {Van Tent}, {Vibert},
  {Vielva}, {Villa}, {Vittorio}, {Wand elt}, {Wehus}, {White}, {White},
  {Zacchei}, \& {Zonca}}]{2018arXiv180706209P}
{Planck Collaboration}, {Aghanim}, N., {Akrami}, Y., {et~al.} 2018, arXiv
  e-prints, arXiv:1807.06209

\bibitem[{{Rahmati} {et~al.}(2013){Rahmati}, {Pawlik}, {Rai{\v{c}}evi{\'c}}, \&
  {Schaye}}]{2013MNRAS.430.2427R}
{Rahmati}, A., {Pawlik}, A.~H., {Rai{\v{c}}evi{\'c}}, M., \& {Schaye}, J. 2013,
  \mnras, 430, 2427

\bibitem[{{Rahmati} \& {Schaye}(2018)}]{2018MNRAS.478.5123R}
{Rahmati}, A., \& {Schaye}, J. 2018, \mnras, 478, 5123

\bibitem[{{Robertson} {et~al.}(2015){Robertson}, {Ellis}, {Furlanetto}, \&
  {Dunlop}}]{2015ApJ...802L..19R}
{Robertson}, B.~E., {Ellis}, R.~S., {Furlanetto}, S.~R., \& {Dunlop}, J.~S.
  2015, \apjl, 802, L19

\bibitem[{{Schaye}(2001)}]{2001ApJ...559..507S}
{Schaye}, J. 2001, \apj, 559, 507

\bibitem[{{Schenker} {et~al.}(2012){Schenker}, {Stark}, {Ellis}, {Robertson},
  {Dunlop}, {McLure}, {Kneib}, \& {Richard}}]{schenker12}
{Schenker}, M.~A., {Stark}, D.~P., {Ellis}, R.~S., {et~al.} 2012, \apj, 744,
  179

\bibitem[{{Shapiro} \& {Giroux}(1987)}]{1987ApJ...321L.107S}
{Shapiro}, P.~R., \& {Giroux}, M.~L. 1987, \apjl, 321, L107

\bibitem[{{Shapiro} {et~al.}(2004){Shapiro}, {Iliev}, \&
  {Raga}}]{2004MNRAS.348..753S}
{Shapiro}, P.~R., {Iliev}, I.~T., \& {Raga}, A.~C. 2004, \mnras, 348, 753

\bibitem[{{Shull} {et~al.}(2012){Shull}, {Harness}, {Trenti}, \&
  {Smith}}]{2012ApJ...747..100S}
{Shull}, J.~M., {Harness}, A., {Trenti}, M., \& {Smith}, B.~D. 2012, \apj, 747,
  100

\bibitem[{{Sirko}(2005)}]{2005ApJ...634..728S}
{Sirko}, E. 2005, \apj, 634, 728

\bibitem[{{So} {et~al.}(2014){So}, {Norman}, {Reynolds}, \&
  {Wise}}]{2014ApJ...789..149S}
{So}, G.~C., {Norman}, M.~L., {Reynolds}, D.~R., \& {Wise}, J.~H. 2014, \apj,
  789, 149

\bibitem[{{Sobacchi} \& {Mesinger}(2014)}]{2014MNRAS.440.1662S}
{Sobacchi}, E., \& {Mesinger}, A. 2014, \mnras, 440, 1662

\bibitem[{{Trac} \& {Cen}(2007)}]{2007ApJ...671....1T}
{Trac}, H., \& {Cen}, R. 2007, \apj, 671, 1

\bibitem[{{Trac} {et~al.}(2008){Trac}, {Cen}, \& {Loeb}}]{2008ApJ...689L..81T}
{Trac}, H., {Cen}, R., \& {Loeb}, A. 2008, \apjl, 689, L81

\bibitem[{{Trac} \& {Pen}(2004)}]{2004NewA....9..443T}
{Trac}, H., \& {Pen}, U.-L. 2004, \na, 9, 443

\bibitem[{{Viel} {et~al.}(2013){Viel}, {Becker}, {Bolton}, \&
  {Haehnelt}}]{2013PhRvD..88d3502V}
{Viel}, M., {Becker}, G.~D., {Bolton}, J.~S., \& {Haehnelt}, M.~G. 2013, \prd,
  88, 043502

\bibitem[{{Wagner} {et~al.}(2015){Wagner}, {Schmidt}, {Chiang}, \&
  {Komatsu}}]{2015MNRAS.448L..11W}
{Wagner}, C., {Schmidt}, F., {Chiang}, C.~T., \& {Komatsu}, E. 2015, \mnras,
  448, L11

\bibitem[{{Wang} {et~al.}(2019){Wang}, {Yang}, {Fan}, {Wu}, {Yue}, {Li},
  {Bian}, {Jiang}, {Ba{\~n}ados}, {Schindler}, {Findlay}, {Davies}, {Decarli},
  {Farina}, {Green}, {Hennawi}, {Huang}, {Mazzuccheli}, {McGreer}, {Venemans},
  {Walter}, {Dye}, {Lyke}, {Myers}, \& {Haze Nunez}}]{2019ApJ...884...30W}
{Wang}, F., {Yang}, J., {Fan}, X., {et~al.} 2019, \apj, 884, 30

\bibitem[{{Worseck} {et~al.}(2014){Worseck}, {Prochaska}, {O'Meara}, {Becker},
  {Ellison}, {Lopez}, {Meiksin}, {M{\'e}nard}, {Murphy}, \&
  {Fumagalli}}]{2014MNRAS.445.1745W}
{Worseck}, G., {Prochaska}, J.~X., {O'Meara}, J.~M., {et~al.} 2014, \mnras,
  445, 1745

\bibitem[{{Wu} {et~al.}(2019){Wu}, {McQuinn}, {Kannan}, {D'Aloisio}, {Bird},
  {Marinacci}, {Dav{\'e}}, \& {Hernquist}}]{2019MNRAS.490.3177W}
{Wu}, X., {McQuinn}, M., {Kannan}, R., {et~al.} 2019, \mnras, 490, 3177

\end{thebibliography}

\begin{appendix}

\section{Numerical convergence and other tests}
\label{sec:convergence}
\begin{figure}
\includegraphics[width=8.5cm]{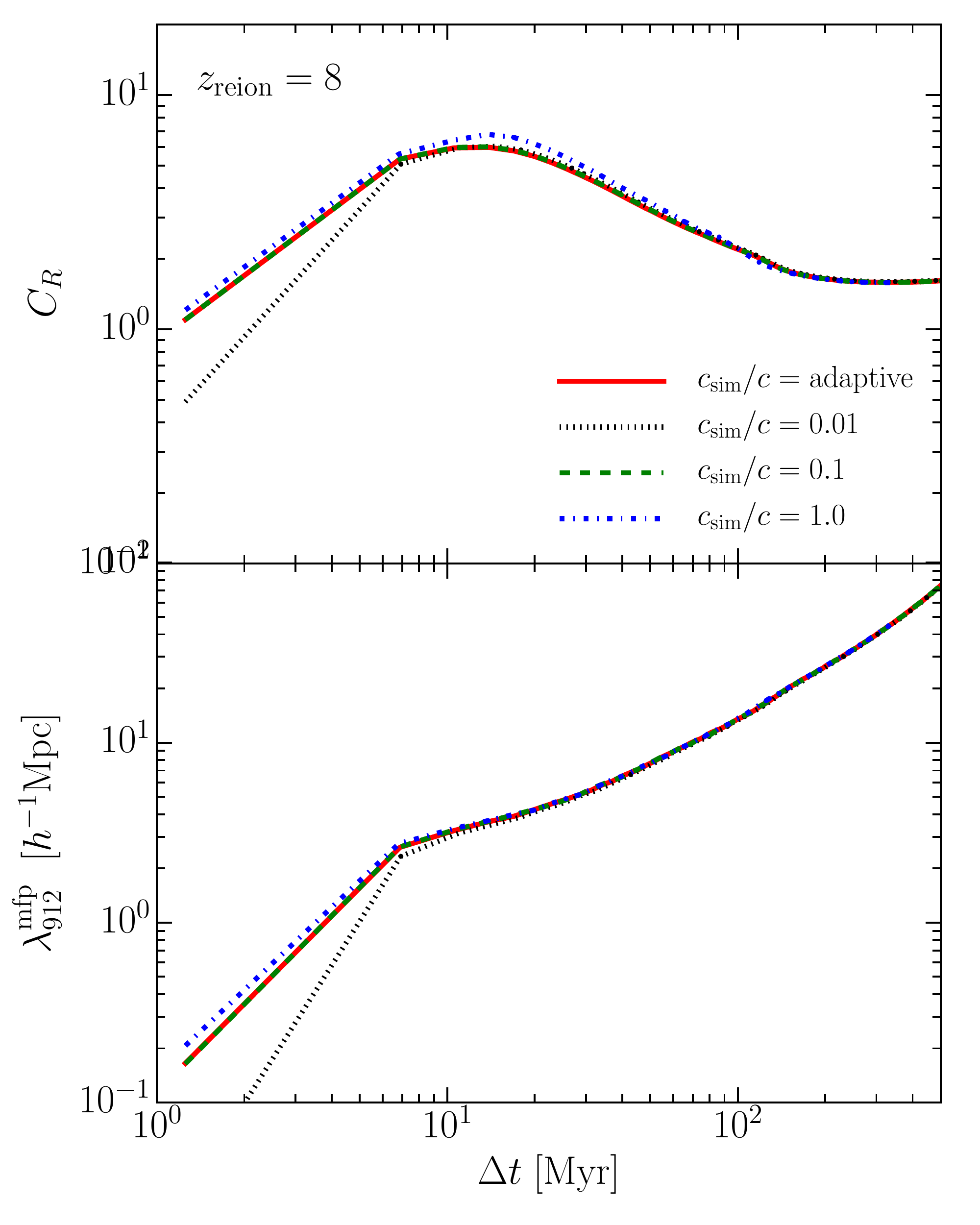}
\includegraphics[width=8.5cm]{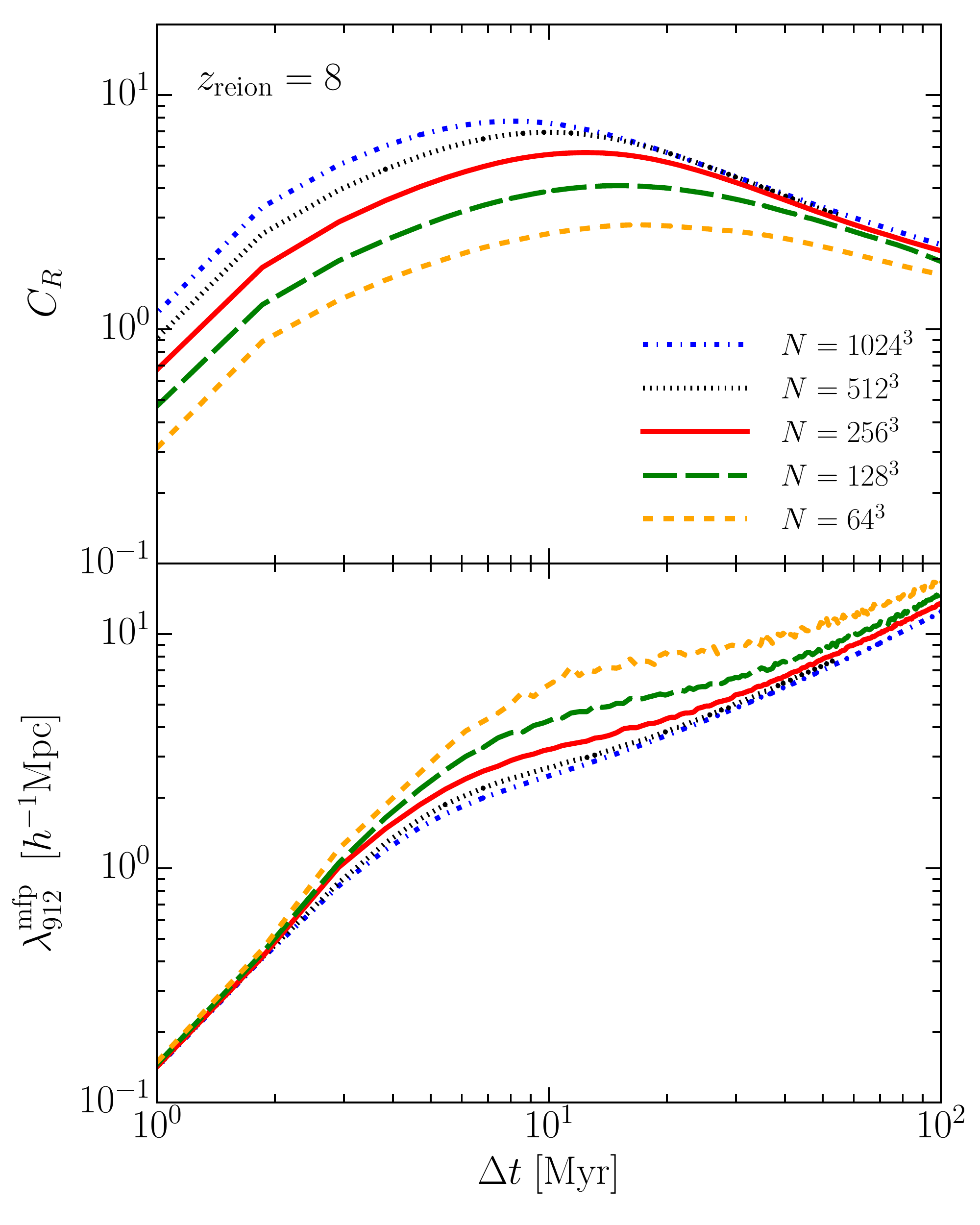} 
\caption{Numerical convergence tests of our radiation-hydrodynamics simulations. { Top and bottom panels show the clumping factor and mean free path, respectively.}  All test simulations shown here use $L_{\rm box} = 256 h^{-1}$ kpc (with 1/8th the volume of our production runs), $\Gamma_{-12}=0.3$, and $\zreion=8$.  {\it Left:} test of our adaptive reduce speed of light approximation.  The solid/red curve employs our adaptive method, while the other line-styles correspond to constant light speeds.  The blue/dot-dashed curve shows a full speed of light run with $c_{\rm sim} = c$. All test runs here have $N=256^3$.  Our adaptive method is accurate to better than $10$\%. {\it Right:} test of convergence with respect to grid size, $N$.  The red curve has the same resolution as our production runs.  Our production runs are well-converged after $\Delta t = 10$ Myr.  Before this time, the { clumping factor  in the $N=256^3$ run deviates from that of the $N=1024^3$ run by a factor of $1.7$, which is similar to the level of uncertainty introduced by X-ray pre-heating of the gas. On the other hand, the mean free path always agrees within $\approx 30\%$}.  }
\label{fig:convergence}
\end{figure}

To reduce computational costs, our simulations adopt an adaptive reduced speed of light approximation as described in \S \ref{sec:methods}.  We have tested this approximation against a series of runs with different light speeds in boxes with $L_{\rm box} = 256 h^{-1}$ kpc, $N=256^3$, $\zreion=8$, and $\Gamma_{-12}=0.3$.  In order to keep the RT domain sizes equal to those of our production runs, we have set $N_{\rm dom} = 8^3$ for all runs with $L_{\rm box} = 256 h^{-1}$ kpc.     The left panel of Fig. \ref{fig:convergence} shows the { clumping factor ($C_R$; top) and mean free path ($\MFP$; bottom) from our speed of light tests.}  In the plot legend, we quote the adopted speed of light in units of $c$.  The red/solid curve with the ``adaptive" label corresponds to the algorithm used for our production runs. We find that, while the $c_{\rm sim}/c=0.01$ results deviate significantly from the rest at early times ($\Delta < 10$ Myr), { both $C_R$ and $\MFP$ are relatively insensitive to $c_{\rm sim}/c$ at later times}.  This is consistent with our argument in \S \ref{sec:methods} justifying our adaptive $c_{\rm sim}/c$.  In the early phases of the ionization, the relevant velocity scale is that of the I-fronts, which is much smaller than $c_{\rm sim} = 0.1 c$ for the $\Gamma_{-12}$ adopted in this paper (see eq. \ref{eqn:vIF}).  However, once the I-fronts have traversed the domains in our simulations, the relevant scale is the sound speed, which is always very much smaller than $c_{\rm sim} = 0.01 c$.  We find that our adaptive approximation deviates from the $c_{\rm sim}/c=1$ case by at most $\approx 10$\% near the peak of $C_R$.

We have also tested the numerical convergence of our results with respect to $N$ (which, recall, corresponds to the hydro and RT grid sizes, and the dark matter particle number, which are all equal to one another). We used the adaptive speed of light approximation and the same run parameters as in the above paragraph, except that we varied $N$ from $1024^3$ down to $64^3$ by factors of 8. In these tests, $N=256^3$ corresponds to the same resolution as our production runs.  The right panel of Fig. \ref{fig:convergence} shows a comparison of { $C_R$ and $\MFP$}. We find that our production simulations are well converged after $\Delta t = 10$ Myr.  Before $\Delta t = 10$ Myr, however, even our $N=1024^3$ simulation may be approaching convergence at the $\lesssim 10$\% level, suggesting that structures smaller in size than $\Delta x = 10 h^{-1}$ kpc contribute significantly to the recombination rate at these early times.  The $N=256^3$ and $N=1024^3$ results for $C_R$ deviate by a factor of $1.7$ for $\Delta t < 10$ Myr.  We note that this is similar to the difference we found between our fidicual runs and our pre-heating run with $T_i=1,000$ K (which differ by a factor of $1.9$).  In other words, our production runs -- in which the gas evolves adiabatically until the ionizing radiation turns on -- fail to converge at a similar level to the uncertainties introduced by X-ray pre-heating of the gas.  { On the other hand, the bottom-right panel of Fig. \ref{fig:convergence} shows that mean free paths agree within $\approx 30 \%$ between the $N=256^3$ and $N=1024^3$ runs.}   

\begin{figure}
\includegraphics[width=8.5cm]{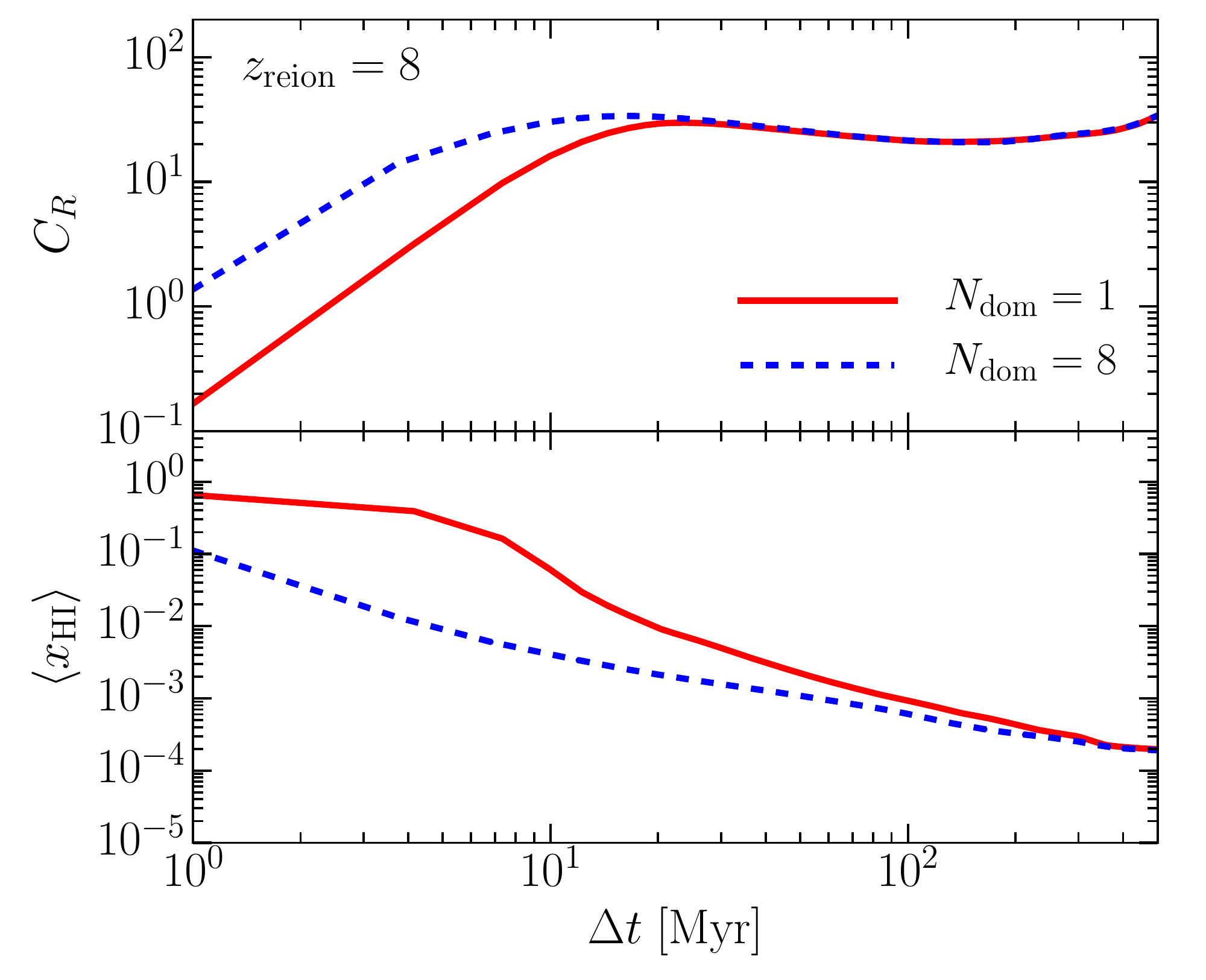}
\hspace{0.5cm}
\includegraphics[width=8.5cm]{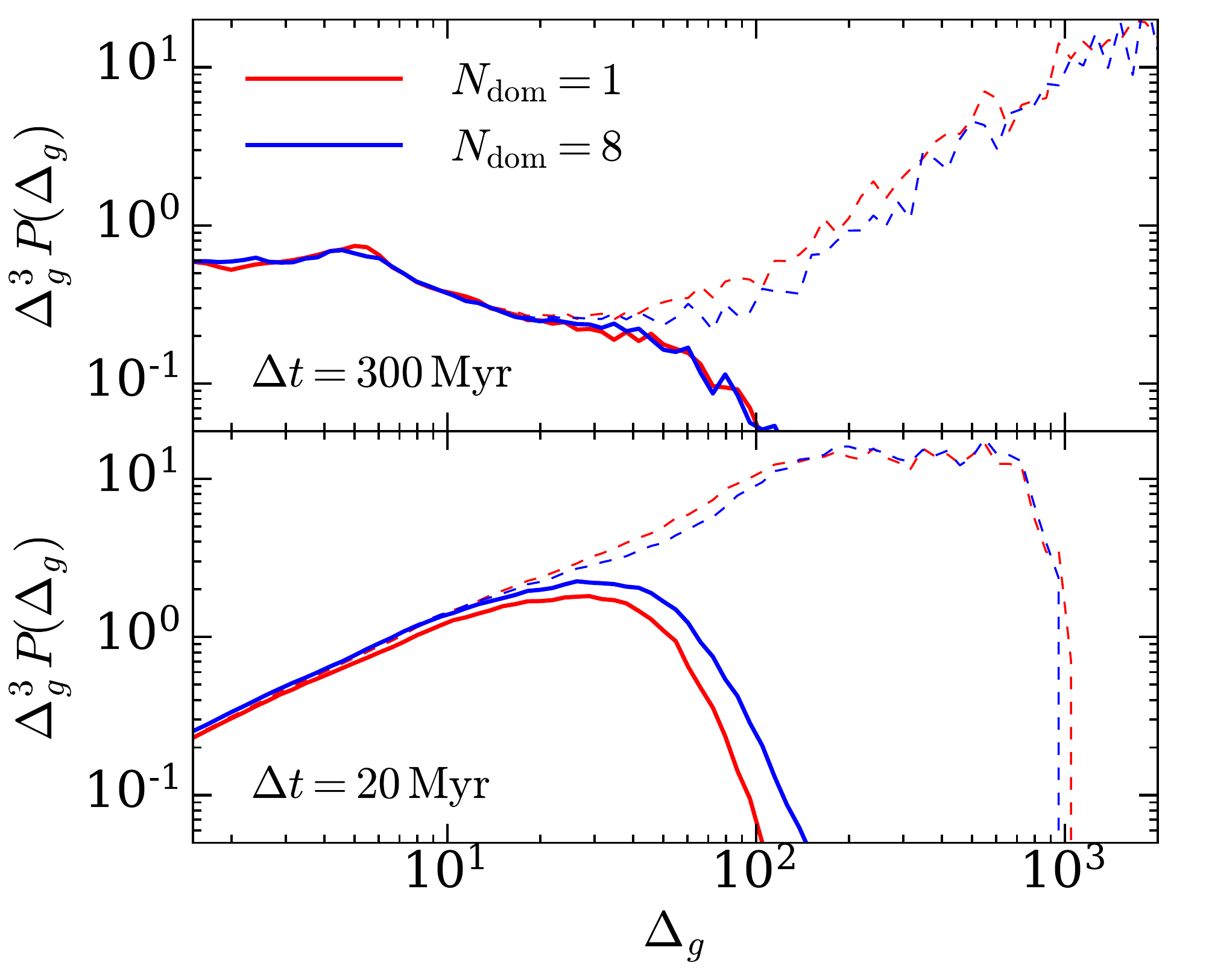} 
\caption{A test for potential artifacts from our domain RT structure.  We show results from two simulations with $L_{\rm box}=256 h^{-1}$ kpc, $\zreion = 8$, $\Gamma_{-12}=0.3$, and $\delta/\sigma = \sqrt{3}$.  The DC mode is used to increase the amount of dense peaks in the box, which increases the potential for spurious effects arising from the domain structure.  We compare two runs with $N_{\rm dom = }=1$ and 8, respectively. {\it Left:} comparison of the clumping factors (top) and volume-weighted mean neutral fractions (bottom) from the two runs. The I-fronts in the $N_{\rm dom}=1$ simulation take much longer to traverse the domains, illustrating the need for small domains to capture the rapid early evolution of $C_R$ in the un-relaxed gas.  By $\Delta t = 20$ Myr the I-fronts have made it through the $N_{\rm box}=1$ box and the two runs agree. {\it Right:} Comparison of the gas density distributions for ionized (solid) and all gas (dashed).  The top and bottom panels show two snapshops in time. Agreement between the distributions suggests that the domains do not introduce significant spurious effects from ionized over-dense gas that would otherwise have remained neutral.  }
\label{fig:convergenceB}
\end{figure}

As described in \S \ref{sec:methods}, our RT method employs a domain structure that is designed to clarify the interpretation of our results by making $\zreion$ and $\Gamma_{-12}$ as uniform as possible throughout the simulation box.  We have tested the robustness of our domain method.  We compare results from two test runs, each with $L_{\rm box} = 256 h^{-1}$ kpc and $N=256^3$. The primary goal here is to test whether the ionization of over-dense regions (which otherwise might have remained neutral) by the domain structure introduces spurious effects in our results, e.g. artificially boosting $C_R$.  To this end, we employ a DC mode with $\delta /\sigma = \sqrt{3}$ in order to enhance the abundance of over-densities in the box.  The first run uses one domain, i.e. the ``domain" is the entire $L_{\rm box}= 256 h^{-1}$ kpc box, and the other uses $N_{\rm dom } = 8^3$ domains of size $L_{\rm dom} = 32 h^{-1}$ kpc -- the same size as in our production runs.  The first case is equivalent to using $N_{\rm dom} = 4^3$ domains of size $L_{\rm dom}= 256 h^{-1}$ kpc in our production boxes.  The left panel of Fig. \ref{fig:convergenceB} compares $C_R$ (top) and the volume-weighted average neutral fraction, $\langle x_{\rm HI}\rangle$ (bottom).  The crucial difference between the runs is that the neutral fractions reach 1\% at $\Delta t = 20$ Myr and $\Delta t = 5$ Myr for $N_{\rm dom}=1$ and $N_{\rm dom}=8$, respectively. As a result, the $N_{\rm dom}=1$ run fails to capture the rapid evolution of $C_R$ at $\Delta t < 20$ Myr because the I-fronts are still moving through the larger domains at these times. The $C_R$ agree quite well once I-fronts have traversed both boxes.  The right-hand panels of Fig. \ref{fig:convergenceB} compare gas density distributions at $\Delta t= 20$ Myr and $\Delta t = 300$ Myr, respectively. The solid curves correspond to ionized gas, while the dashed curves correspond to the full gas distribution (see \S \ref{sec:ionPDF} for a definition of $\Delta^3_g P(\Delta_g)$).  We find good agreement between the $N_{\rm dom}=1$ and $N_{\rm dom}=8$ distributions, even at $\Delta t = 20$ Myr, in spite of the neutral fractions differing by a factor of 5.  These results suggest that the domains used in our production runs do not introduce significant spurious effects at high densities.

\section{Averaging over DC modes with Gauss-Hermite Quadrature}
\label{DCmodes}

Here we present our Gauss-Hermite Quadrature method for averaging results from DC-mode simulations. Consider a physical quantity, $Y(t)$. (For concreteness, one may consider the clumping factor -- see \S \ref{sec:recrate}).   Imagine dividing up the universe into cubic sub-volumes equal to our simulation volumes and measuring $Y$ in each.  If the mean density contrasts were sufficient to describe evolution in each sub-volume, we could compute the average of $Y$ by performing the integral $\langle Y(t) \rangle = \int_{-1}^{\infty} d \delta_{\rm NL}~Y(\delta_{\rm NL},t)~P_V(\delta_{\rm NL},t)$, where $\delta_{\rm NL}(t)$ is the mean non-linear density contrast of a given sub-volume, and $P_V$ is the volume-weighted probability distribution of $\delta_{\rm NL}$.  This assumes that the evolution within a volume is entirely characterized by $\delta_{\rm NL}(t)$, ignoring correlations with other box-scale quantities (the most important of which is likely the tidal field).  Because the box-scale overdensity controls growth, this assumption is likely accurate for the $Y$ of interest -- the clumping factor and the mean free path.

We can recast this integral in terms of $\delta$, the linearly extrapolated initial density contrasts, by writing $\langle Y(t) \rangle = \int_{-\infty}^{\infty} d \delta~Y(\delta,t)~P_V(\delta,t)$. Here, $P_V(\delta,t) = P_{\rm L}(\delta) \cdot V_{\rm E}/V_{\rm L}(\delta,t)$, where $P_L(\delta)$ is the Gaussian probability distribution of $\delta$, and $V_{\rm E}/V_{\rm L}(\delta,t)$ is the fraction of the final volume that is occupied by Lagrangian elements with initial density $\delta$.  The latter accounts for the contraction/expansion of Lagrangian volume elements from gravitational collapse.  Appealing to the spherical collapse model, we approximate this fraction as $V_{\rm E}/V_{\rm L}(\delta,t) = \eta / (1+\delta_{\rm NL}(\delta,t))$,where $\eta$ is a ``fudge factor" that accounts for deviations in volume conservation from approximating the universe as a collection of spherically collapsing/expanding regions.  Under these approximations, the average becomes     

\begin{equation}
\left< Y(t) \right> \approx \eta \int_{-\infty}^{\infty} d \delta~\frac{Y(\delta,t)}{ 1+\delta_{\rm NL}(\delta,t)} \frac{1}{ \sqrt{2 \pi \sigma^2}}\exp\left(\frac{-\delta^2}{2 \sigma^2} \right).
\label{eq:genGI}
\end{equation}
We fix $\eta$ by demanding that the integral is normalized to unity in the absence of $Y(\delta,t)$ in the integrand. Equation (\ref{eq:genGI}) is of course approximate, but we find that $\eta$ is never far from unity in our calculations, with a maximum deviation of $1-\eta = 6\%$ at $z=5.5$ (the lowest redshift in our DC mode simulations). 

In practice, we must evaluate the integral (\ref{eq:genGI}) using measurements of $Y(\delta,t)$ from just a handful of simulations. The optimal values of $\delta/\sigma$ can be determined by appealing to Gaussian quadrature. Under the change of variable, $x = \delta/\sqrt{2 \sigma^2}$, we can recast eq. (\ref{eq:genGI}) to the form 

\begin{equation}
\langle Y(t) \rangle = \frac{\eta}{\sqrt{\pi}}\int_{-\infty}^{\infty} d x ~f(x ,t) \exp\left(-x^2\right),
\end{equation}
where we have defined $f(x,t) = Y(x ,t)/( 1+\delta_{\rm NL}(x,t) )$.  To this integral we can readily apply the technique of Gauss-Hermite quadrature,
\begin{equation}
\langle Y(t) \rangle \approx \frac{\eta}{\sqrt{\pi}} \sum_{i=1}^{n} w_i~f(x_i,t),
\label{eq:quadrature}
\end{equation}
where $n$ is the number of sample points, $w_i$ are the Gauss-Hermite weights, and $x_i$ are the roots of the Hermite polynomial $H_i(x)$. Note that if $f$ were a polynomial up to degree five, then $\langle Y \rangle$ could be computed exactly using just $n=3$ sample points.  In what follows, we assume that $f$ can be well-approximated by a polynomial of degree five, as we expect the clumping factor and mean free path to be smooth functions of the DC mode.  We use 3 sample points determined by the roots of $H_3(x)$ to be $\delta/\sigma = 0, \pm \sqrt{3}$.  We therefore complement our mean-density runs with two DC mode simulations with $\delta/\sigma =  \pm \sqrt{3}$ (see Table \ref{tab:sims}).  In this case, the weights are $w_1 = 2\sqrt{\pi}/3 $ and $w_2 = w_3 =  \sqrt{\pi}/6$, where $w_1$ corresponds to the $\delta/\sigma = 0$ evaluation.  

We apply this method to compute the mean clumping factor, $\langle C_R(t) \rangle$, and mean free path, $\langle \MFP(t) \rangle$.  For the latter we take $Y(t)$ to be the absorption coefficient of the box, $\kappa(t) \equiv 1 / \MFP(t)$.  We then take $\langle \MFP(t) \rangle \approx 1/\langle \kappa(t) \rangle$, which is equivalent to assuming that correlations in $\kappa$ between neighboring $L_{\rm box} = 1 h^{-1}$ Mpc patches of the universe are negligible.

\end{appendix}

\end{document}